%% file: Bsmumu_LHC_Autumn13_Paper.tex
\documentclass[12pt,a4paper]{article}

\usepackage{amsmath}

\usepackage{xspace}

\usepackage{epsfig}

\usepackage{rotating}

\usepackage{dcolumn}

\usepackage{url}
\usepackage{changepage}

\usepackage{ifthen} 
\usepackage{subfigure} 
\usepackage{rotating}

\usepackage{graphicx}  

\usepackage{xspace}  
\usepackage{color}
\usepackage{colortbl}

\newboolean{pdflatex}
\setboolean{pdflatex}{true} 
\newboolean{articletitles}
\setboolean{articletitles}{true} 

\newboolean{uprightparticles}
\setboolean{uprightparticles}{false} 
\usepackage{amssymb}
\usepackage{amsfonts}
\usepackage{upgreek}
\usepackage{rotating}
\usepackage{multirow}
\usepackage{xspace}
\usepackage{color}
\usepackage{amsmath}
\usepackage{graphicx}
\usepackage{bm}
\usepackage{framed}
\usepackage{comment}
\usepackage{ctable}
\usepackage{placeins}
\usepackage{graphicx}
\usepackage{color}
\usepackage{colortbl}
\usepackage{lineno}
\usepackage{cuted}
\usepackage{float}
\usepackage{caption}
\usepackage{hyperref}

\hypersetup{%
  colorlinks=false, linktocpage=false, pdfborder={0 0 0}, pdfstartpage=1, 
 pdfstartview=FitV,%
}

\usepackage[superscript,nomove]{cite}
\usepackage{mciteplus}

\input{lhcb-symbols-def}

\input{our-symbols-def}
\input{numbers}

\renewcommand{\thefootnote}{\fnsymbol{footnote}}
\floatstyle{plain}
\newfloat{extended}{thp}{lop}
\DeclareCaptionLabelSeparator*{pipe}{ \boldmath$\vert$ }
\floatname{extended}{\textbf{Extended Data Figure}}
\newlength{\OneColumnWidth}
\newlength{\TwoColumnWidth}
\newlength{\HalfColumnWidth}
\begin{document}

\begin{titlepage}

\belowpdfbookmark{Title page}{title}

\includecomment{toProduceBBL}

\pagenumbering{roman}

\vspace*{-1.5cm}

\centerline{\large EUROPEAN ORGANIZATION FOR NUCLEAR RESEARCH (CERN)}

\vspace*{1.5cm}

\hspace*{-5mm}\begin{tabular*}{16cm}{lc@{\extracolsep{\fill}}r}
&& CMS-BPH-13-007 \\
&& LHCb-PAPER-2014-049 \\
&& CERN-PH-EP-2014-220 \\
&& May 13,2015 \\
\end{tabular*}

\vspace*{5cm}

\noindent{\bf\Large\boldmath{
Observation of the rare $\phantom{\Big(}\!\!\Bsmm$  decay from the combined analysis of CMS and LHCb data}}

\vspace*{5cm}

\begin{center}

\normalsize{The CMS and LHCb Collaborations\footnote{Lists of the participants and their affiliations appear at the end of the Letter.}}

\end{center}

\vspace{\fill}



\vspace*{1.cm}

\vspace{\fill}

\end{titlepage}

\cleardoublepage

\renewcommand{\thefootnote}{\arabic{footnote}}
\setcounter{footnote}{0}

\pagestyle{plain} 
\setcounter{page}{1}
\pagenumbering{arabic}

\input{content_Paper}

\end{document}

%% file: lhcb-symbols-def.tex
\def\lhcb {\mbox{LHCb}\xspace}

\ifthenelse{\boolean{uprightparticles}}%
{

 \def\Pmu         {\ensuremath{\upmu}\xspace}                 
 \def\Pnu         {\ensuremath{\upnu}\xspace}

 \def\PDelta      {\ensuremath{\Delta}\xspace}                 
 \def\PXi      {\ensuremath{\Xi}\xspace}                 
 \def\PLambda      {\ensuremath{\Lambda}\xspace}                 
 \def\PSigma      {\ensuremath{\Sigma}\xspace}                 
 \def\POmega      {\ensuremath{\Omega}\xspace}                 
 \def\PUpsilon      {\ensuremath{\Upsilon}\xspace}                 
 
 \def\PB      {\ensuremath{\mathrm{B}}\xspace}                 
                  
 \def\PD      {\ensuremath{\mathrm{D}}\xspace}

 \def\PK      {\ensuremath{\mathrm{K}}\xspace}

 \def\PZ      {\ensuremath{\mathrm{Z}}\xspace}                 
                  
 \def\Pb      {\ensuremath{\mathrm{b}}\xspace}

 \def\Pi      {\ensuremath{\mathrm{i}}\xspace}

 \def\Ps      {\ensuremath{\mathrm{s}}\xspace}

}
{

 \def\Pmu         {\ensuremath{\mu}\xspace}                 
 \def\Pnu         {\ensuremath{\nu}\xspace}

 \mathchardef\PDelta="7101
 \mathchardef\PXi="7104
 \mathchardef\PLambda="7103
 \mathchardef\PSigma="7106
 \mathchardef\POmega="710A
 \mathchardef\PUpsilon="7107
                  
 \def\PB      {\ensuremath{B}\xspace}                 
                  
 \def\PD      {\ensuremath{D}\xspace}

 \def\PK      {\ensuremath{K}\xspace}

 \def\PZ      {\ensuremath{Z}\xspace}                 
                  
 \def\Pb      {\ensuremath{b}\xspace}

 \def\Pi      {\ensuremath{i}\xspace}

 \def\Ps      {\ensuremath{s}\xspace}

}

\def\mup        {\ensuremath{\Pmu^+}\xspace}
\def\mun        {\ensuremath{\Pmu^-}\xspace} 
\def\mumu       {\ensuremath{\Pmu^+\Pmu^-}\xspace}

\def\Z      {\ensuremath{\PZ}\xspace}

\def\squark    {\ensuremath{\Ps}\xspace}

\def\bquark    {\ensuremath{\Pb}\xspace}

  \def\Kbar  {\kern 0.2em\overline{\kern -0.2em \PK}{}\xspace}

  \def\Dbar    {\kern 0.2em\overline{\kern -0.2em \PD}{}\xspace}

\def\B       {\ensuremath{\PB}\xspace}
\def\Bbar    {\ensuremath{\kern 0.18em\overline{\kern -0.18em \PB}{}}\xspace}

\def\Bu      {\ensuremath{\B^+}\xspace}

\def\Bd      {\ensuremath{\B^0}\xspace}
\def\Bs      {\ensuremath{\B^0_\squark}\xspace}
\def\Bsb     {\ensuremath{\Bbar^0_\squark}\xspace}
\def\Bdb     {\ensuremath{\Bbar^0}\xspace}

  \def\Y#1S{\ensuremath{\PUpsilon{(#1S)}}\xspace}

\def\Lbar {\ensuremath{\kern 0.1em\overline{\kern -0.1em\PLambda}}\xspace}

\newcommand{\decay}[2]{\ensuremath{#1\!\to #2}\xspace}         

\def\to                 {\ensuremath{\rightarrow}\xspace}

\def\AT#1     {\ensuremath{A_{\mathrm{T}}^{#1}}\xspace}           

\def\Bsmm     {\decay{\Bs}{\mup\mun}}

\def\C#1      {\ensuremath{\mathcal{C}_{#1}}\xspace}                       
\def\Cp#1     {\ensuremath{\mathcal{C}_{#1}^{'}}\xspace}                    
\def\Ceff#1   {\ensuremath{\mathcal{C}_{#1}^{\mathrm{(eff)}}}\xspace}        
\def\Cpeff#1  {\ensuremath{\mathcal{C}_{#1}^{'\mathrm{(eff)}}}\xspace}       
\def\Ope#1    {\ensuremath{\mathcal{O}_{#1}}\xspace}                       
\def\Opep#1   {\ensuremath{\mathcal{O}_{#1}^{'}}\xspace}                    

\newcommand{\unit}[1]{\ensuremath{\rm\,#1}\xspace}          

\newcommand{\tev}{\ensuremath{\mathrm{\,Te\kern -0.1em V}}\xspace}
\newcommand{\gev}{\ensuremath{\mathrm{\,Ge\kern -0.1em V}}\xspace}
\newcommand{\mev}{\ensuremath{\mathrm{\,Me\kern -0.1em V}}\xspace}
\newcommand{\kev}{\ensuremath{\mathrm{\,ke\kern -0.1em V}}\xspace}
\newcommand{\ev}{\ensuremath{\mathrm{\,e\kern -0.1em V}}\xspace}
\newcommand{\gevc}{\ensuremath{{\mathrm{\,Ge\kern -0.1em V\!/}c}}\xspace}
\newcommand{\mevc}{\ensuremath{{\mathrm{\,Me\kern -0.1em V\!/}c}}\xspace}
\newcommand{\tevcc}{\ensuremath{{\mathrm{\,Te\kern -0.1em V\!/}c^2}}\xspace}
\newcommand{\gevcc}{\ensuremath{{\mathrm{\,Ge\kern -0.1em V\!/}c^2}}\xspace}
\newcommand{\gevgevcccc}{\ensuremath{{\mathrm{\,Ge\kern -0.1em V^2\!/}c^4}}\xspace}
\newcommand{\mevcc}{\ensuremath{{\mathrm{\,Me\kern -0.1em V\!/}c^2}}\xspace}

\def\m    {\ensuremath{\rm \,m}\xspace}

\def\invfb   {\ensuremath{\mbox{\,fb}^{-1}}\xspace}

\def\gsim{{~\raise.15em\hbox{$>$}\kern-.85em
          \lower.35em\hbox{$\sim$}~}\xspace}
\def\lsim{{~\raise.15em\hbox{$<$}\kern-.85em
          \lower.35em\hbox{$\sim$}~}\xspace}

\def\pt         {\mbox{$p_{\rm T}$}\xspace}

\def\tell1  {TELL1\xspace}
\def\ukl1   {UKL1\xspace}



%% file: our-symbols-def.tex
\def\bujpsik{\ensuremath{B^+ \to J/\psi K^+}\xspace}
\def\bujpsikmm{\ensuremath{B^+ \to J/\psi (\mumu) K^+}\xspace}
\def\bdkpi{\ensuremath{\Bd \to K^+ \pi^-}\xspace}

\newcommand{\BF}{branching fraction\xspace}
\newcommand{\BFs}{branching fractions\xspace}

\newcommand{\Bq}{\ensuremath{B^0_{(s)}}\xspace}

\newcommand{\Bsmumu}{\ensuremath{\Bs\to\mu^+\mu^-}\xspace}
\newcommand{\Bdmumu}{\ensuremath{\Bd\to\mu^+\mu^-}\xspace}
\def\bsmumu{\Bsmumu}
\def\bdmumu{\Bdmumu}
\newcommand{\Bsdmumu}{\ensuremath{B_{(s)}^0\to\mu^+\mu^-}\xspace}
\newcommand{\Bsd}{\ensuremath{B_{(s)}^0}\xspace}

\newcommand{\bhhprime}{\ensuremath{B^0_{(s)}\to h^+h^{\prime-}}\xspace}

\newcommand{\BuJpsiK}{\ensuremath{B^+\to J/\psi K^+}\xspace}

\newcommand{\Bqmumu}{\ensuremath{\Bq\to\mu^+\mu^-}\xspace}

\newcommand{\BsJpsiPhi}{\ensuremath{B^0_s\to J/\psi \phi}\xspace}
\def\bdpimunu{\ensuremath{B^0 \to \pi^- \mu^+ \nu}\xspace}
\def\bskmunu{\ensuremath{B^0_s \to K^- \mu^+ \nu}\xspace}
\def\Panu{\overline \Pnu}
\def\lbpmunu{\ensuremath{\Lambda^0_b \to p \mu^- \Panu}\xspace}

\def\bpiZeromumu{\ensuremath{B^{0} \to \pi^{0} \mu^+ \mu^-}\xspace}
\def\bpiPlusmumu{\ensuremath{B^{+} \to \pi^{+} \mu^+ \mu^-}\xspace}

\newcommand{\BRof}[1]{\ensuremath{{\cal B}(#1)}\xspace}

\newcommand{\fdfsinline}{\ensuremath{f_d/\kern -0.1emf_s}\xspace}
\newcommand{\fdfsinlinecompact}{\ensuremath{^{f_d}\kern -0.2em/\kern -0.22em_{f_s}}\xspace}
\newcommand{\fdfs}{\fdfsinline}

\newcommand{\fdfuinline}{\ensuremath{f_d/\kern -0.1emf_u}\xspace}

\newcommand{\mmumu}{\ensuremath{m_{\mu^+\mu^-}}\xspace}

\newcommand{\tblue}[1]{\textcolor{black}{#1}}

\def\SBd{\ensuremath{\mathcal{S}^{\Bd}_{\rm{SM}}}\xspace}
\def\SBs{\ensuremath{\mathcal{S}^{B^0_s}_{\rm{SM}}}\xspace}
\def\SBq{\ensuremath{\mathcal{S}^{B^0_{(s)}}_{\rm{SM}}}\xspace}
\def\RB{\ensuremath{{\mathcal{R}}}\xspace}

\def\sm{SM\xspace}

%% file: numbers.tex
\newcommand{\BRBsmm}{\ensuremath{3.66\pm0.23}\xspace}
\newcommand{\BRBdmm}{\ensuremath{1.06\pm0.09}\xspace}
\newcommand{\RBSM}{\ensuremath{0.0295^{+0.0028}_{-0.0025}}\xspace} 

\newcommand{\eN}[1]{\ensuremath{#1 \times 10^{-9}}\xspace}
\newcommand{\eT}[1]{\ensuremath{#1 \times 10^{-10}}\xspace}

\def\LHCbBsmumuMeasureMod{\ensuremath{\left( 2.7 \,^{+1.1}_{-0.9} \right) \times 10^{-9}}\xspace}
\def\LHCbBdmumuMeasureMod{\ensuremath{\left( 3.3 \,^{+2.4}_{-2.1} \right) \times 10^{-10}}\xspace}

\def\CMSBsmumuMeasureMod{\ensuremath{\left( 2.8 \,^{+1.0}_{-0.9} \right) \times 10^{-9}}\xspace}
\def\CMSBdmumuMeasureMod{\ensuremath{\left( 4.4 \,^{+2.2}_{-1.9} \right) \times 10^{-10}}\xspace}

\def\BsmumuMeasure{\ensuremath{\left( 2.8 \,^{+0.7}_{-0.6} \right) \times 10^{-9}}\xspace}
\def\BdmumuMeasure{\ensuremath{\left( 3.9 \,^{+1.6}_{-1.4} \right) \times 10^{-10}}\xspace}

\def\BsmumuSignificance{\ensuremath{6.2}\xspace}
\def\BdmumuSignificance{\ensuremath{3.2}\xspace}

\def\BsmumuSignalStrenght{\ensuremath{0.76 \,^{+0.20}_{-0.18}}\xspace}
\def\BdmumuSignalStrenght{\ensuremath{3.7 \,^{+1.6}_{-1.4}}\xspace}

\def\BsmumuDistanceToSM{\ensuremath{1.2}\xspace}
\def\BdmumuDistanceToSM{\ensuremath{2.2}\xspace}

\def\RatioBdoverBs{\ensuremath{ 0.14 \,^{+0.08}_{-0.06} }\xspace}

\def\RatioDistanceToSM{\ensuremath{2.3}\xspace}

\def\BdmumuFCOneSigmaInt{\ensuremath{[2.5, 5.6] \times 10^{-10}}\xspace} 
\def\BdmumuFCTwoSigmaInt{\ensuremath{[1.4, 7.4] \times 10^{-10}}\xspace} 

\def\BdmumuFCSignificance{\ensuremath{3.0}\xspace} 

%% file: content_Paper.tex
\input{executive_summary}
\input{introduction_reorg}

\input{results}

\input{discussion}

\renewcommand\refname{\bf\normalsize References}
\bibliographystyle{LHCb}
\nocite{Giles:1984yg,Avery:1987cv,Avery:1989qi,Ammar:1993ez,Bergfeld:2000ui} 
\nocite{Albrecht:1987rj} 
\nocite{Albajar:1988iq,Albajar:1991ct} 
\nocite{Abe:1996et,Abe:1998ah,Acosta:2004xj,Abulencia:2005pw,Aaltonen:2011fi,Aaltonen:2013as} 
\nocite{Acciarri:1996us} 
\nocite{Abbott:1998hc,Abazov:2004dj,Abazov:2007iy,Abazov:2010fs,Abazov:2013wjb} 
\nocite{Chang:2003yy} 
\nocite{Aubert:2004gm,Aubert:2007hb} 
\nocite{Aaij:2011rja,Aaij:2012ac,LHCb:2011ac,Aaij:2012nna,Aaij:2013aka} 
\nocite{Chatrchyan:2011kr,Chatrchyan:2012rga,Chatrchyan:2013bka} 
\nocite{Aad:2012pn} 
\nocite{LHCb-CONF-2013-011,DeBruyn:2012wk,Khodjamirian:2011jp,LHCB-PAPER-2014-003,Higgs-Comb}

\makeatletter
\renewcommand\@biblabel[1]{#1.}
%
\input{ref1}

\input{end_notes}
\clearpage

\input{methods_extended}

%
\input{ref2}

\clearpage
%
\input{CMS_AL.tex}
\input{LHCb_AL.tex}
\clearpage
%
\input{extended_data}
\clearpage

%% file: executive_summary.tex
{\bf\boldmath The standard model of particle physics describes the fundamental particles 
and their interactions via the strong, electromagnetic, and weak forces.                   
It provides precise predictions for measurable quantities that can be tested 
experimentally.                                                                                          
The probabilities, or branching fractions, of the strange $B$ meson (\Bs) and the \Bd meson decaying into 
two oppositely charged muons ($\mu^+$ and $\mu^-$) are especially interesting                                         
because of their sensitivity to theories that extend the standard model. 
The standard model predicts that the \Bsmumu and \Bdmumu decays are very rare, with about 
four of the former occurring for every billion \Bs mesons produced and 
one of the latter occurring for every 10 billion \Bd mesons~\cite{Bobeth:2013uxa}.
A difference in the  observed branching fractions with respect to the predictions of the standard model 
would provide a direction in which the standard model should be extended.
Before the Large Hadron Collider (LHC) at CERN~\cite{LHC} started operating,
no evidence for either decay mode had been found.
Upper limits on the branching fractions were an order of magnitude above 
the standard model predictions.
The CMS (Compact Muon Solenoid) and LHCb (Large Hadron Collider beauty)
collaborations have performed a joint analysis of the data from
proton-proton collisions that they collected in 2011 at a centre-of-mass 
energy of seven teraelectronvolts and in 2012 at eight teraelectronvolts. 
Here we report the first observation of the \Bsmumu 
decay, with a statistical significance exceeding six standard deviations, 
and the best measurement so far of its branching fraction. 
Furthermore, we obtained evidence for the \Bdmumu decay with a statistical 
significance of three standard deviations.
Both measurements are statistically compatible with standard model predictions and
allow stringent constraints to be placed on theories beyond the standard model.
The LHC experiments will resume data taking in 2015, recording proton-proton 
collisions at a centre-of-mass energy of 13 teraelectronvolts, which
will approximately double the production rates for \Bs and \Bd 
mesons and lead to further improvements in the precision of these
crucial tests of the standard model.}

%% file: introduction_reorg.tex

Experimental particle physicists have been testing the predictions of the standard model of particle physics (\sm) with increasing precision since the 1970s. 
Theoretical developments have kept pace by improving the accuracy of the \sm predictions as the experimental results gained in precision.
In the course of the past few decades, the \sm has passed critical tests from experiment, 
but it does not address some profound questions about the nature of the Universe. 
For example, the existence of dark matter, which has been confirmed by cosmological data~\cite{Ade:2013zuv}, is not accommodated by the \sm. 
It also fails to explain the origin of the asymmetry between matter and antimatter, which after the Big Bang led 
to the survival of the 
tiny amount of matter currently present in the Universe~\cite{Ade:2013zuv,Gavela}. 
Many theories have been proposed to modify the \sm to provide solutions to these open questions. 

%
%
\newlength{\ThisFigureWidth}
\newlength{\ThisFigureHalfWidth}
\newlength{\ThisFigureItemWidth}
\setlength{\ThisFigureWidth}{130mm}
\setlength{\ThisFigureItemWidth}{30mm}
\begin{figure}[b]
\centering%
\begin{minipage}[b]{\ThisFigureWidth}
\includegraphics[width=\ThisFigureItemWidth]{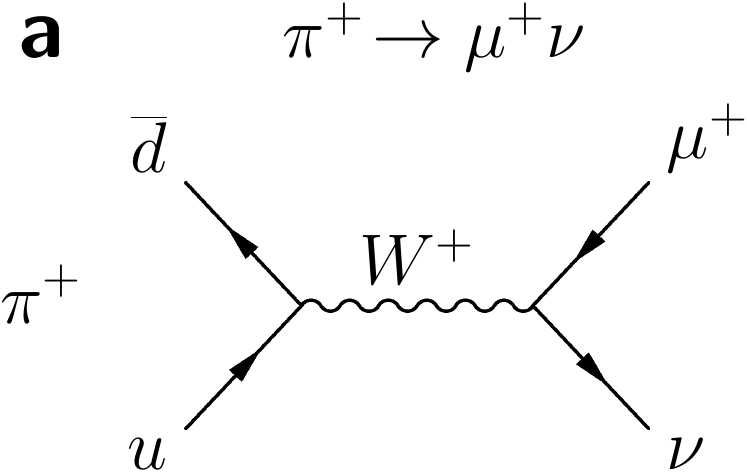}
\hfill
\includegraphics[width=\ThisFigureItemWidth]{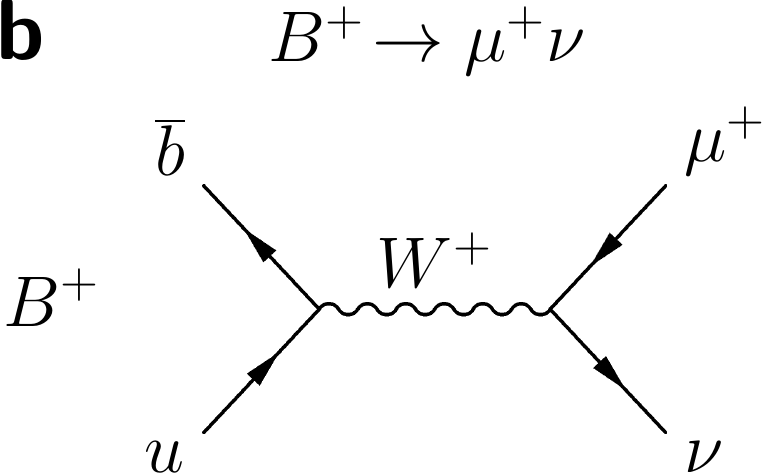}
\hfill
\includegraphics[width=\ThisFigureItemWidth]{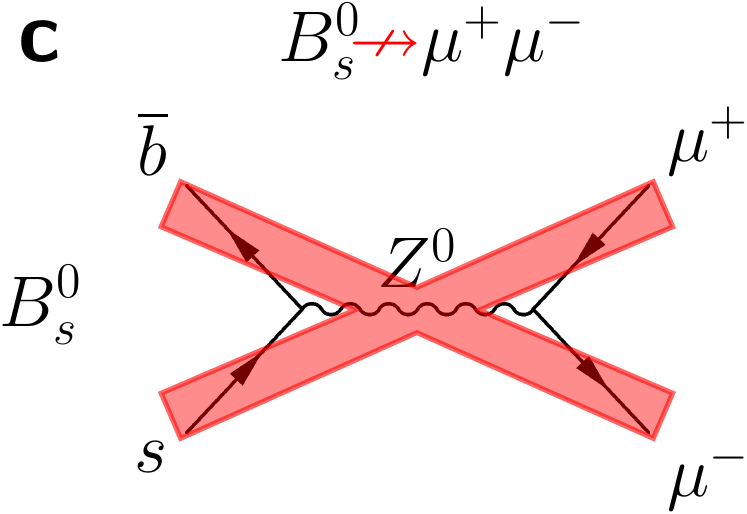}
\hfill
\includegraphics[width=\ThisFigureItemWidth]{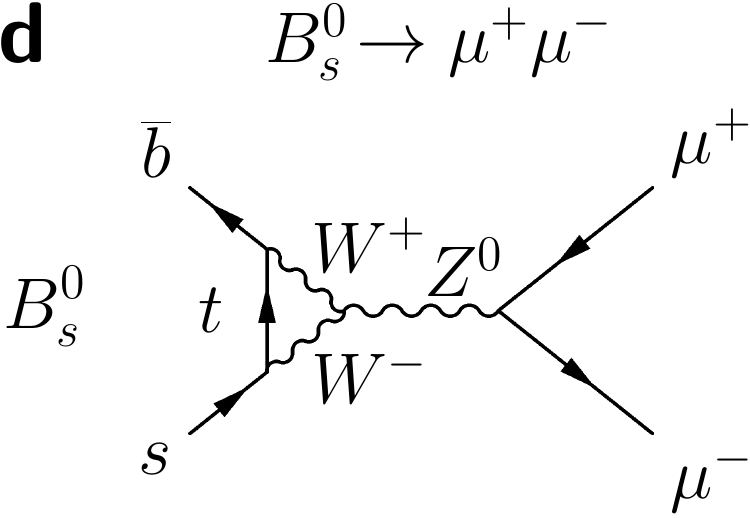}
\end{minipage}

\vspace{7mm}

\setlength{\ThisFigureWidth}{102mm}
\begin{minipage}[b]{\ThisFigureWidth}
\includegraphics[width=\ThisFigureItemWidth]{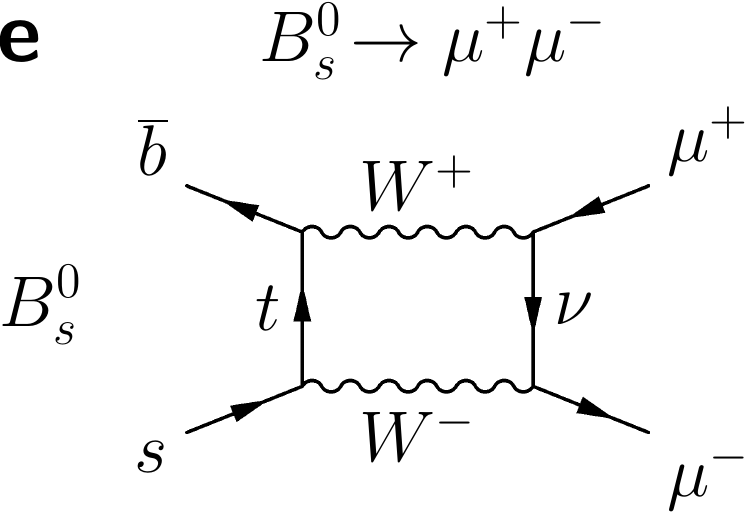}
\hfill
\includegraphics[width=\ThisFigureItemWidth]{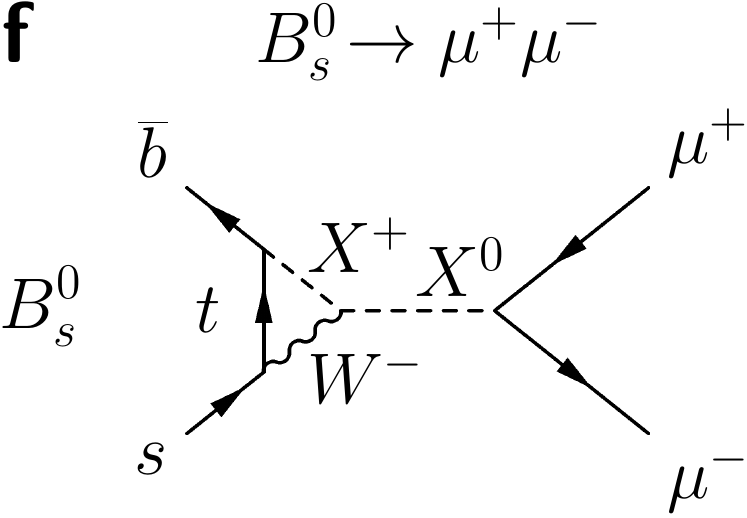}
\hfill
\includegraphics[width=\ThisFigureItemWidth]{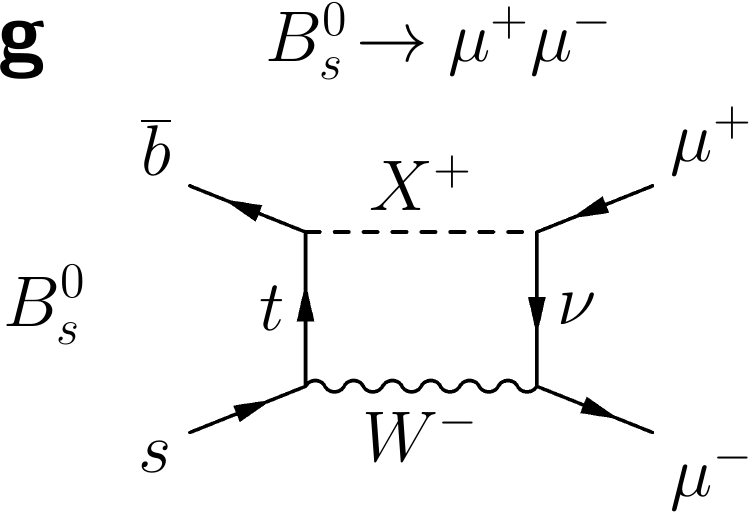}
\end{minipage}
%
\caption{\textbf{\boldmath Feynman diagrams related to the \Bsmumu decay:} 
\textbf{\textsf{a}}, $\pi^{+}$ meson decay through charged-current process;
\textbf{\textsf{b}}, \Bu meson decay through the charged-current process; 
\textbf{\textsf{c}}, a \Bs decay through the direct flavour changing neutral current process, which is forbidden in the \sm, as indicated by the large red ``X; 
\textbf{\textsf{d}} and \textbf{\textsf{e}}, higher-order flavour changing neutral current processes for the \Bsmumu decay allowed in the \sm; and
\textbf{\textsf{f}} and \textbf{\textsf{g}}, examples of processes for the same decay in theories extending the \sm, where new particles, 
denoted as $X^0$ and $X^+$, can alter the decay rate.\label{feynman}}
\end{figure}

The \Bs and \Bd mesons are unstable particles that decay via the weak interaction. The measurement of the branching fractions of the very 
rare decays of these mesons into a dimuon ($\mu^{+}\mu^{-}$) final state is especially interesting.

At the elementary level, the weak force is composed of a `charged current' and a `neutral current' mediated by the $W^{\pm}$
and $Z^0$ bosons, respectively. An example of the charged current is the decay of the $\pi^{+}$ meson,
which consists of an up ($u$) quark of electrical charge $+2/3$ of the charge of the proton and a down ($d$) antiquark of charge $+1/3$. A pictorial representation of this process, 
known as a Feynman diagram, is shown in Fig.~\ref{feynman}a. The $u$ and $d$ quarks are `first generation' or lowest mass quarks.
Whenever a decay mode is specified in this Letter, the charge conjugate mode is implied.

The $\B^{+}$ meson is similar to the $\pi^{+}$, except that the light $d$ antiquark is replaced by the heavy `third generation' (highest mass quarks)
beauty ($b$) antiquark, which has a charge of $+1/3$ and a mass of $\sim$5\gevcc (about five times the mass of a proton). 
The decay $B^{+} \rightarrow \mu^{+}\nu$, represented in Fig.~\ref{feynman}b, is allowed but highly
suppressed because of angular momentum considerations (helicity suppression) and because it involves transitions between
quarks of different generations (CKM suppression), specifically the third and first generations of quarks. 
All $b$ hadrons, including the $B^{+}$, \Bs and \Bd mesons, decay predominantly via the transition 
of the $b$ antiquark to a `second generation' (intermediate mass quarks) charm ($c$) antiquark, which is less CKM suppressed, in final states with charmed hadrons. 
Many allowed decay modes, which typically
involve charmed hadrons and other particles, have angular momentum configurations that are not helicity suppressed. 

The neutral \Bs meson is similar to the $B^{+}$ except that the $u$ quark is replaced by a second generation strange ($s$)
quark of charge $-1/3$. The decay of the \Bs meson to two muons, shown in Fig.~\ref{feynman}c, is forbidden at the elementary level
because the $\Z^0$ cannot couple directly to quarks of different flavours, that is, there are no direct `flavour changing neutral
currents'. However, it is possible to respect this rule and still have this decay occur through the `higher order' transitions such as those shown
in Fig.~\ref{feynman}d and e. These are highly suppressed because each additional interaction vertex reduces their probability
of occurring significantly. They are also helicity and CKM suppressed. 
Consequently, the branching fraction for the \Bsmumu decay is expected to be very small compared to the dominant $b$ antiquark to $c$ antiquark 
transitions. The corresponding decay of the \Bd meson, where a $d$ quark 
replaces the $s$ quark, is even more CKM suppressed because it requires a jump across two quark generations rather than just one.

The \BFs of these two decays, $\mathcal{B}$, accounting for higher-order electromagnetic and strong interaction effects, 
and using lattice quantum chromodynamics to compute the \Bs and \Bd meson decay constants~\cite{Witzel:2013sla,Na:2012kp,Bazavov:2011aa}, 
are reliably calculated~\cite{Bobeth:2013uxa} in the \sm.
Their values are
\mbox{$\BRof{\Bsmumu}_\textrm{SM} = \eN{(\BRBsmm)}$} and
\mbox{$\BRof{\Bdmumu}_\textrm{SM} = \eT{(\BRBdmm)}$}.

Many theories that seek to go beyond the standard model (BSM) include new phenomena and particles~\cite{Huang:1998vb,Choudhury:1998ze}, 
such as in the diagrams shown in Fig.~\ref{feynman}f and g, 
that can significantly modify the \sm \BFs. 
In particular, theories with additional Higgs bosons~\cite{Babu:1999hn,Bobeth:2001sq} predict possible enhancements to the \BFs.
A significant deviation of either of the two branching fraction measurements from the \sm predictions would give insight on 
how the \sm should be extended. 
Alternatively, a measurement compatible with the \sm could provide strong constraints on BSM theories.

The ratio of the branching fractions of the two decay modes provides powerful discrimination among BSM theories~\cite{Buras:2003td}. 
It is predicted in the \sm~\cite{Bobeth:2013uxa,Aoki:2013ldr,PDG2012,HFAG} to be
$\RB \equiv \BRof{\Bdmumu}_\textrm{SM} / \BRof{\Bsmumu}_\textrm{SM} = \RBSM$.
Notably, BSM theories with the property of minimal flavour violation~\cite{MFV_basepaper} 
predict the same value as the \sm for this ratio.

The first evidence for the decay \Bsmumu was presented by the LHCb collaboration in 2012~\cite{Aaij:2012nna}. 
Both CMS and LHCb later published results from all data collected in proton-proton collisions at 
centre-of-mass energies of 7\tev in 2011 and 8\tev in 2012. 
The measurements had comparable precision and were in good agreement~\cite{Chatrchyan:2013bka,LHCb-PAPER-2013-046}, 
although neither of the individual results had sufficient precision to constitute the first definitive observation of the \Bs decay to two muons.

In this Letter, the two sets of data are combined and analysed simultaneously 
to exploit fully the statistical power of the data and to account for the main correlations between them.
The data correspond to total integrated luminosities of $25\invfb$ and $3\invfb$ for the CMS and LHCb experiments, respectively, 
equivalent to a total of approximately $10^{12}$ \Bs and \Bd mesons produced in the two experiments together.
Assuming the branching fractions given by the \sm and accounting for the detection efficiencies, 
the predicted numbers of decays to be observed in the two experiments together are 
about 100 for \Bsmumu and 10 for \Bdmumu.

%
The CMS~\cite{Chatrchyan:2008aa} and LHCb~\cite{Alves:2008zz} detectors are designed to measure \sm phenomena with high precision and search for possible deviations.
The two collaborations use different and complementary strategies.
In addition to performing a broad range of precision tests of the \sm and studying the newly-discovered Higgs boson~\cite{Aad:2012tfa,Chatrchyan:2012ufa}, 
CMS is designed to search for and study new particles with masses from about 100\gevcc to a few\tevcc.
Since many of these new particles would be able to decay into $b$ quarks and many of the \sm measurements also involve $b$ quarks, 
the detection of $b$-hadron decays was a key element in the design of CMS.
The LHCb collaboration has optimised its detector to study matter-antimatter asymmetries and rare decays of particles containing $b$ quarks, 
aiming to detect deviations from precise \sm predictions that would indicate BSM effects.
These different approaches, reflected in the design of the detectors, lead to instrumentation of complementary angular regions with respect to the 
LHC beams, to operation at different proton-proton collision rates, and to selection of $b$ quark events with different efficiency 
(for experimental details, see Methods). 
In general, CMS operates at a higher instantaneous luminosity than LHCb but has a lower efficiency for reconstructing low-mass particles, 
resulting in a similar sensitivity to LHCb for \Bd or \Bs (denoted hereafter \Bsd) mesons decaying into two muons.

Muons do not have strong nuclear interactions and are too massive to emit a significant fraction of their energy by electromagnetic radiation.
This gives them the unique ability to penetrate dense materials, such as steel, and register signals in
detectors embedded deep within them. Both experiments use this characteristic to identify muons.  

%
The experiments follow similar data analysis strategies.
Decays compatible with \Bsdmumu (candidate decays) are found by combining the reconstructed trajectories (tracks) of oppositely charged particles 
identified as muons.
The separation between genuine \Bsdmumu decays and random combinations of two muons (combinatorial background), most often from semi-leptonic decays of
two different $b$ hadrons, is achieved using the dimuon invariant mass, \mmumu, 
and the established characteristics of \Bsd-meson decays.
For example, because of their lifetimes of about 1.5\,ps and their production at the LHC with momenta between a few \gevc and $\sim100\gevc$, 
\Bsd mesons travel up to a few centimetres before they decay.
Therefore, the \Bsdmumu `decay vertex', from which the muons originate, is required to be displaced with respect to the `production vertex',
the point where the two protons collide.
Furthermore, the negative of the \Bsd candidate's momentum vector is required to point back to the production vertex. 

These criteria, amongst others that have some ability to distinguish known signal events from background events, are combined 
into boosted decision trees (BDT)~\cite{Breiman,Adaboost,Hocker:2007ht}.
A BDT is an ensemble of decision trees each placing different selection requirements on the individual variables to achieve the best 
discrimination between `signal-like' and  `background-like' events. 
Both experiments evaluated many variables for their discriminating power and each chose the best set of about ten
to be used in its respective BDT. These include variables related to the quality of the reconstructed tracks of the muons;
kinematic variables such as transverse momentum (with respect to the beam axis) of the individual muons and of the
\Bsd candidate; variables related to the decay vertex topology and fit quality, such as candidate decay length; and isolation variables, 
which measure the activity in terms of other particles in the vicinity of the two muons or their displaced vertex.  
A BDT must be `trained'  on collections of known background and signal events to generate the selection requirements on the variables and 
the weights for each tree. In the case of CMS, the background events
used in the training are taken from intervals of dimuon mass above and below the signal region in data, while simulated events are used for the signal.
The data are divided into disjoint sub-samples and the BDT trained on one sub-sample is applied to a different sub-sample to avoid any bias. 
LHCb uses simulated events for background and signal in the training of its BDT.
After training, the relevant BDT is applied to each event in the data, returning a single value for the event, with high values being more signal-like.
To avoid possible biases, both experiments 
kept the small mass interval that includes both the \Bs and \Bd signals blind until all selection criteria were established.  

In addition to the combinatorial background, specific $b$-hadron decays, such as \bdpimunu where the neutrino cannot be detected and the charged pion is
misidentified as a muon, or \bpiZeromumu, where the neutral pion in the decay is not reconstructed, can mimic the dimuon decay of the \Bsd mesons.
The invariant mass of the reconstructed dimuon candidate for these processes (semi-leptonic background) is usually smaller than the mass of the \Bs 
or \Bd meson because the neutrino or another particle is not detected.
There is also a background component from hadronic two-body \Bsd decays (peaking background) as \decay{\Bd}{K^+\pi^-}, when both hadrons from the decay are misidentified as muons. 
These misidentified decays can produce
peaks in the dimuon invariant-mass spectrum near the expected signal, especially for the \Bdmumu decay.
Particle identification algorithms are used to minimise the probability that pions and kaons are misidentified as muons, and thus suppress these background sources. 
Excellent mass resolution is mandatory for distinguishing between \Bd and \Bs mesons with a mass difference of about 87\mevcc and for separating them from backgrounds. 
The mass resolution for \Bsmumu decays in CMS ranges from 32 to 75\mevcc, depending on the direction of the muons relative to the beam axis, 
while LHCb achieves a uniform mass resolution of about 25\mevcc.

The CMS and LHCb data are combined by fitting a common value for each \BF to the data from both experiments. 
The branching fractions are determined from
the observed numbers, efficiency-corrected, of \Bsd mesons that decay into two muons and the total
numbers of \Bsd mesons produced.
Both experiments derive the latter from the number of observed  \bujpsik decays,  whose \BF has been precisely measured elsewhere~\cite{PDG2012}.
Assuming equal rates for  \Bu and \Bd production, this gives the normalisation for \Bdmumu. 
To derive the number of
\Bs mesons from this \Bu decay mode, the ratio of $b$ quarks that form (hadronise into) \Bu mesons to those that form \Bs 
mesons is also needed.
Measurements of this ratio~\cite{LHCb-PAPER-2011-018,LHCb-PAPER-2012-037}, for which there is additional discussion in Methods,
and of the branching fraction \BRof\bujpsik are used to normalise both sets of data and are
constrained within Gaussian uncertainties in the fit.
The use of these two results by both CMS and LHCb is the only significant source of correlation between their individual
branching fraction measurements.  
The combined fit takes advantage of the larger data sample to increase the precision while properly accounting for the correlation.

%
%
\begin{figure}[t]
\centering
\includegraphics[height=70mm]{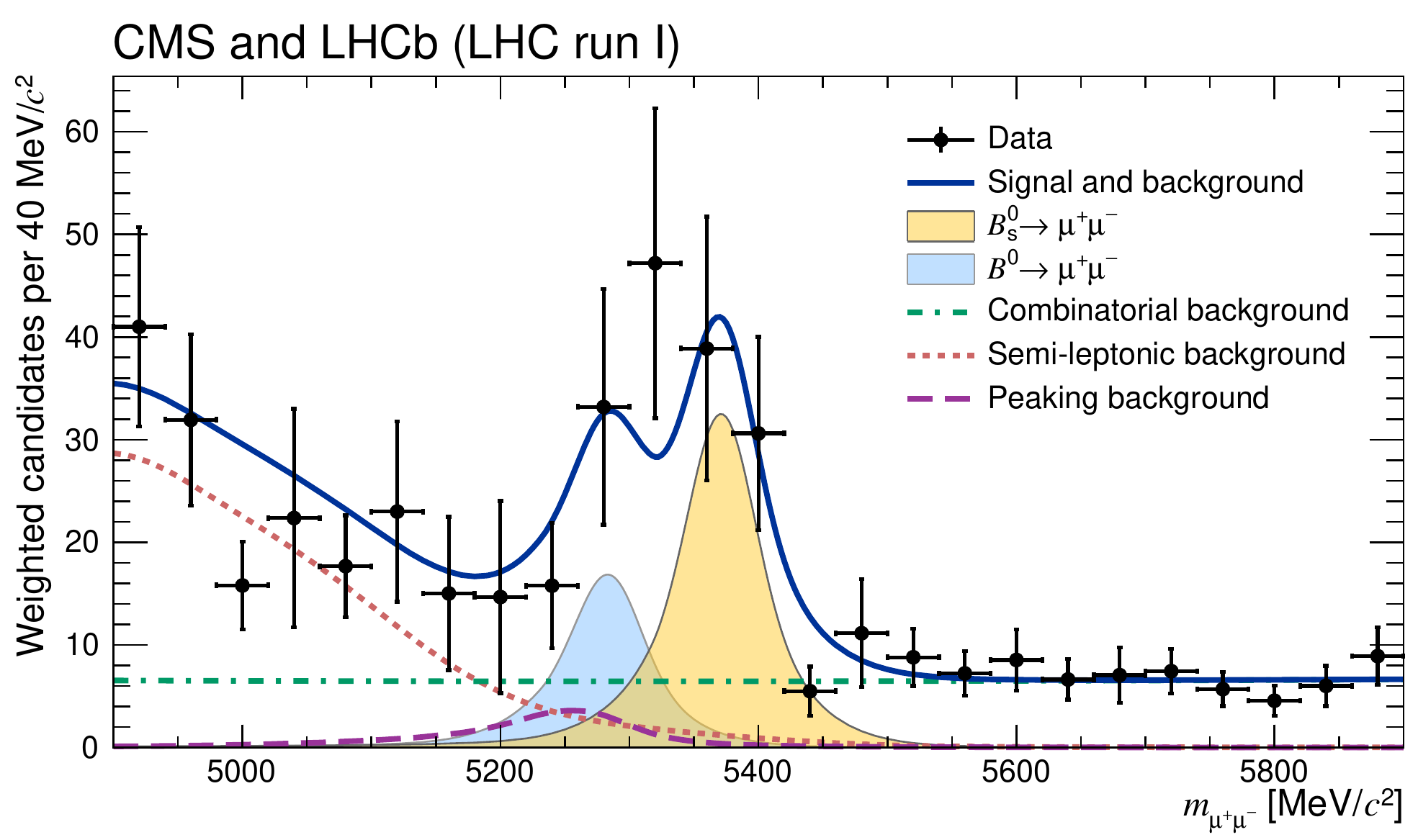}
%
\caption{\textbf{\boldmath Weighted distribution of the dimuon invariant mass, $m_{\mu^+\mu^-}$, for all categories.}
Superimposed on the data points in black are the combined fit (solid blue line) and its components: 
the \Bs (yellow shaded area) and \Bd (light-blue shaded area) signal components; the combinatorial background (dash-dotted green line); 
the sum of the semi-leptonic backgrounds (dotted salmon line); and the peaking backgrounds (dashed violet line).
The horizontal bar on each histogram point denotes the size of the binning, while the vertical bar denotes the 68\% confidence interval.
See main text for details on the weighting procedure.\label{fig:MassWeighted}}
\end{figure}

In the simultaneous fit to both the CMS and LHCb data, the branching fractions of the two signal channels are common parameters of interest and are free to vary.
Other parameters in the fit are considered as nuisance parameters.
Those for which additional knowledge is available are constrained to be near their estimated values by using Gaussian penalties 
with their estimated uncertainties while the others are free to float in the fit. 
The ratio of the hadronisation probability into \Bu and \Bs mesons and the branching fraction of the normalisation channel \BuJpsiK are common, constrained parameters. 
Candidate decays are categorised according to whether they were detected in CMS or LHCb and to the value of the relevant BDT discriminant.
In the case of CMS, they are further categorised according to the data-taking period, and, because of the large variation in mass resolution
with angle, whether the muons are both produced 
at large angles relative to the proton beams (central-region) or at least one muon is emitted at small angle relative to the beams (forward-region).
An unbinned extended maximum likelihood fit to the dimuon invariant-mass distribution, in a region of about $\pm500$\mevcc around the \Bs mass, 
is performed simultaneously in all categories (12 categories from CMS and eight from LHCb).
Likelihood contours in the plane of the parameters of interest,
\BRof\Bdmumu versus \BRof\Bsmumu, are obtained by constructing the
test statistic $-2\Delta\textrm{ln}L$ from the difference in log-likelihood ($\textrm{ln}L$) values
between fits with fixed values for the parameters of interest and the
nominal fit. 
For each of the two branching fractions, a one-dimensional profile
likelihood scan is likewise obtained by fixing only the
single parameter of interest and allowing the other to vary during the
fits. 
Additional fits are performed where the parameters under consideration are the ratio of the branching fractions 
relative to their \sm predictions, $\SBq \equiv  \BRof{\Bqmumu} / \BRof{\Bqmumu}_{\rm{SM}}$, 
or the ratio \RB of the two branching fractions.

%% file: results.tex

The combined fit result is shown for all 20 categories in Extended Data Fig.~\ref{fig:AllBins}.
To represent the result of the fit in a single dimuon invariant-mass spectrum, the mass distributions of all categories, 
weighted according to values of $\rm S / (S+B)$, where S is the expected number of \Bs signals and B is the number of background 
events under the \Bs peak in that category, are added together and shown in Fig.~\ref{fig:MassWeighted}.
The result of the simultaneous fit is overlaid.
An alternative representation of the fit to the dimuon invariant-mass distribution for the six categories with the highest $\rm S / (S+B)$ value 
for CMS and LHCb, as well as displays of events  with high probability to be genuine signal decays, are shown in the 
Extended Data Figs.~\ref{fig:MassBestBins}--\ref{fig:LHCb_event_display}.

%
%
%

\begin{figure}[t]
\centering
\includegraphics[height=55mm]{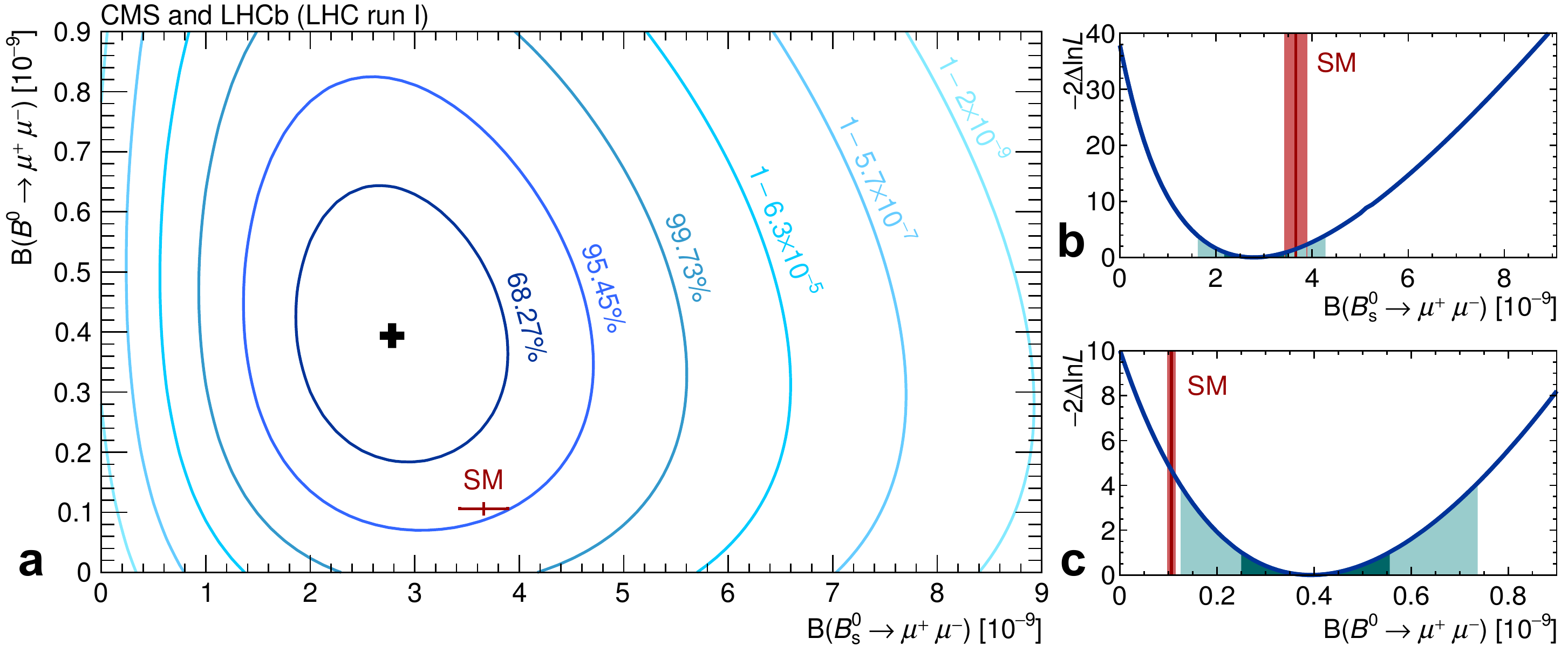}
%
\caption{\textbf{\boldmath Likelihood contours in the $\BRof\Bdmumu$ versus $\BRof\Bsmumu$ plane.} 
The (black) cross in \textbf{\textsf{a}} marks the best-fit central value. 
The \sm expectation and its uncertainty is shown as the (red) marker.
Each contour encloses a region approximately corresponding to the reported confidence level.
\textbf{\textsf{b}}, \textbf{\textsf{c}}, Variations of the test statistic $-2\Delta\textrm{ln}L$ for 
$\BRof\Bsmumu$ (\textbf{\textsf{b}}) and $\BRof\Bdmumu$ (\textbf{\textsf{c}}). 
The dark and light (cyan) areas define the $\pm1\sigma$ and $\pm2\sigma$ confidence intervals for the branching fraction, respectively.
The \sm prediction and its uncertainty for each branching fraction is denoted with the vertical (red) band. 
\label{fig:2d_contour_flavio}}
\end{figure}

The combined fit leads to the measurements
%
$\BRof{\Bsmumu} = \BsmumuMeasure$ and
$\BRof{\Bdmumu} = \BdmumuMeasure$, 
where the uncertainties include both statistical and systematic sources, 
the latter contributing  35\% and 18\% of the total uncertainty for the \Bs and \Bd signals, respectively. 
Using Wilks' theorem~\cite{Wilks}, the statistical significance in unit of standard deviations, $\sigma$, is computed to be 
$\BsmumuSignificance$ for the \Bsmumu decay mode and $\BdmumuSignificance$  for the \Bdmumu mode.
For each signal the null hypothesis that is used to compute the significance includes all background components predicted by the \sm
as well as  the other signal, whose branching fraction is allowed to vary freely.  
The median expected significances assuming the \sm branching fractions are $7.4\,\sigma$ and $0.8\,\sigma$ for the \Bs and \Bd modes, respectively.
Likelihood contours for \BRof{\Bdmumu} versus \BRof{\Bsmumu} are shown in Fig.~\ref{fig:2d_contour_flavio}.
One-dimensional likelihood scans for both decay modes are displayed in the same figure.
In addition to the likelihood scan, the statistical significance and confidence intervals for the \Bd branching fractions are determined using 
simulated experiments. 
This determination yields a significance of $\BdmumuFCSignificance\,\sigma$ for a \Bd signal with respect to the same null hypothesis described above. 
Following the Feldman--Cousins~\cite{Feldman:1997qc} procedure, $\pm1\,\sigma$ and $\pm2\,\sigma$ confidence
intervals for \BRof\Bdmumu of \BdmumuFCOneSigmaInt and \BdmumuFCTwoSigmaInt are obtained, respectively (see Extended Data Fig.~\ref{fig:fc_results_lhcb}).

\begin{figure}[t]
\centering
\includegraphics[width=100mm]{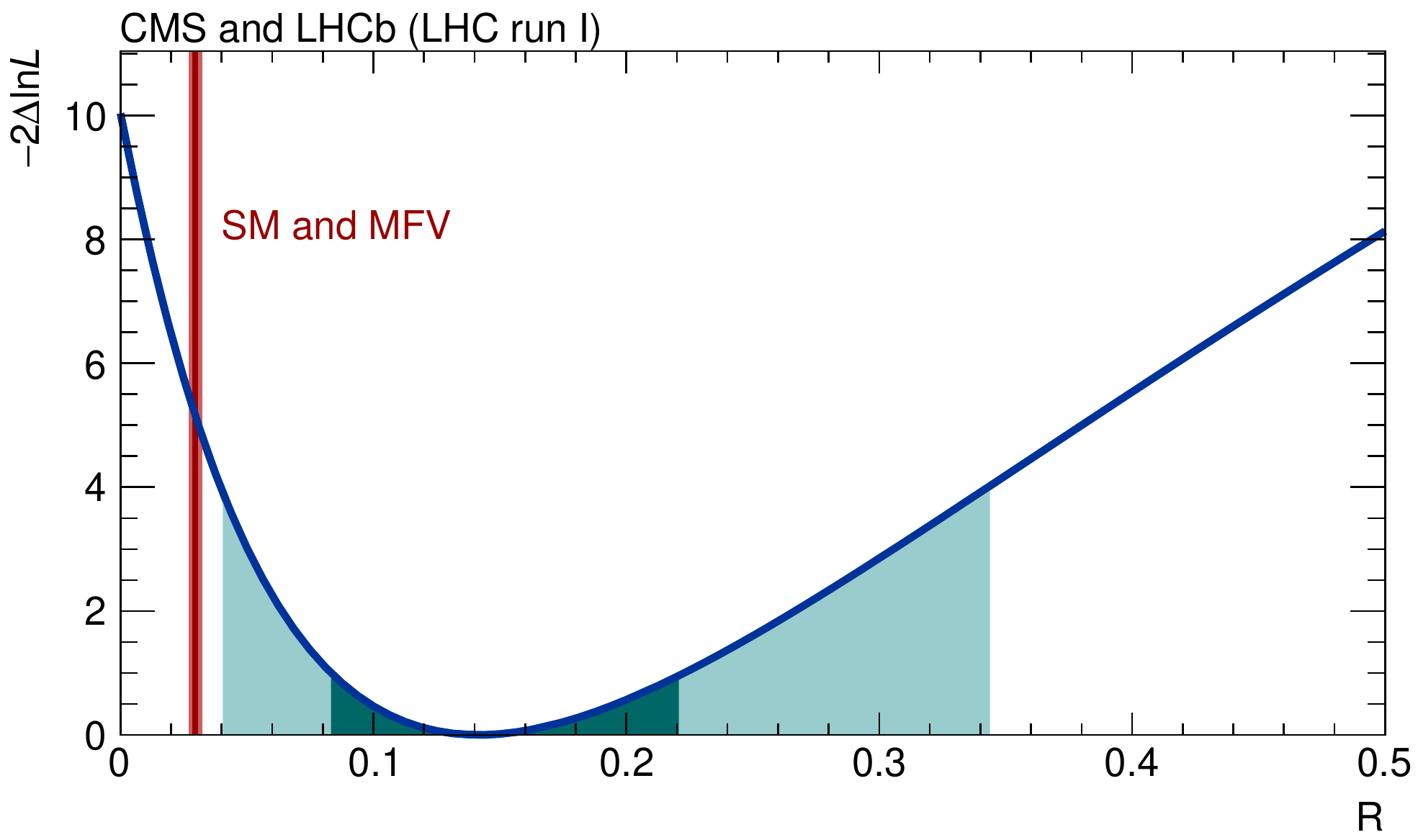}
%
\caption{\textbf{\boldmath Variation of the test statistic $-2\Delta\textrm{ln}L$ as a function of the ratio of branching fractions $\RB \equiv \BRof{\Bdmumu} / \BRof{\Bsmumu}$.} 
The dark and light (cyan) areas define the $\pm1\sigma$ and $\pm2\sigma$ confidence intervals for \RB, respectively.
The value and uncertainty for \RB predicted in the \sm, which is the same in BSM theories with the minimal flavour violation (MFV) property, is denoted with the vertical (red) band. 
\label{fig:1d_R_flavio}}
\end{figure}

The fit for the ratios of the branching fractions relative to their \sm predictions yields
\mbox{\SBs = \BsmumuSignalStrenght} and 
\mbox{\SBd = \BdmumuSignalStrenght}.
Associated likelihood contours and one-dimensional likelihood scans are shown in the Extended Data Fig.~\ref{fig:BFoverSM}.
The measurements are compatible with the \sm branching fractions of the \bsmumu  and \bdmumu decays 
at the $\BsmumuDistanceToSM\,\sigma$  and $\BdmumuDistanceToSM\,\sigma$ level, respectively, when computed from the one-dimensional hypothesis tests.
Finally, the fit for the ratio of branching fractions yields 
\mbox{\RB = \RatioBdoverBs}, 
which is compatible with the \sm at the $\RatioDistanceToSM\,\sigma$ level.
The one-dimensional likelihood scan for this parameter is shown in Fig.~\ref{fig:1d_R_flavio}. 

%
%

%% file: discussion.tex
%
The combined analysis of data from CMS and LHCb, taking  advantage of their full statistical power,
establishes conclusively the existence 
of the \Bsmumu decay and provides an improved  measurement of its branching fraction.
This concludes a search that started more than three decades ago (see Extended Data Fig.~\ref{fig:History}), 
and initiates a phase of precision measurements of the properties of this decay.
It also produces a three standard deviation evidence for the \Bdmumu decay.
The measured branching fractions of both decays are compatible with \sm predictions. 
This is the first time that the CMS and LHCb collaborations have performed a combined analysis of sets of their data 
in order to obtain a  statistically significant observation.

%% file: ref1.tex
\ifx\mcitethebibliography\mciteundefinedmacro
\PackageError{LHCb.bst}{mciteplus.sty has not been loaded}
{This bibstyle requires the use of the mciteplus package.}\fi
\providecommand{\href}[2]{#2}

%% file: end_notes.tex
\paragraph{Acknowledgements} 
We express our gratitude to our colleagues in the CERN
accelerator departments for the excellent performance of the LHC. 
We thank the technical and administrative staff at CERN, at the CMS
institutes and at the LHCb institutes. 
In addition, we gratefully acknowledge the computing centres and personnel of the Worldwide LHC Computing Grid
for delivering so effectively the computing infrastructure essential to our analyses. 
Finally, we acknowledge the enduring support for the construction and operation of the LHC, the CMS
and the LHCb detectors provided by CERN and by many funding agencies.
The following agencies provide support for both CMS and LHCb:
CAPES, CNPq, FAPERJ and FINEP (Brazil);
NSFC (China);
CNRS/IN2P3 (France);
BMBF, DFG, and HGF (Germany);
SFI (Ireland); 
INFN (Italy);
NASU (Ukraine);
STFC (UK);
NSF (USA).
Agencies that provide support for CMS only are: 
BMWFW and FWF (Austria); 
FNRS and FWO (Belgium);
FAPESP (Brazil);
MES (Bulgaria);
CAS and MoST (China);
COLCIENCIAS (Colombia); 
MSES and CSF (Croatia); 
RPF (Cyprus);
MoER, ERC IUT and ERDF (Estonia); 
Academy of Finland, MEC, and HIP (Finland);
CEA (France);
GSRT (Greece); 
OTKA and NIH (Hungary); 
DAE and DST (India); 
IPM (Iran);
NRF and WCU (Republic of Korea); 
LAS (Lithuania); 
MOE and UM (Malaysia); 
CINVESTAV, CONACYT, SEP, and UASLP-FAI (Mexico); 
MBIE (New Zealand); 
PAEC (Pakistan);
MSHE and NSC (Poland);
FCT (Portugal);
JINR (Dubna); 
MON, RosAtom, RAS and RFBR (Russia);
MESTD (Serbia);
SEIDI and CPAN (Spain); 
Swiss Funding Agencies (Switzerland);
MST (Taipei); 
ThEPCenter, IPST, STAR and NSTDA (Thailand); 
TUBITAK and TAEK (Turkey);
SFFR (Ukraine);
DOE (USA).
Agencies that provide support for LHCb only are: 
FINEP (Brazil);
MPG (Germany); 
FOM and NWO (The Netherlands);
MNiSW and NCN (Poland); 
MEN/IFA (Romania); 
MinES and FANO (Russia);
MinECo (Spain);
SNSF and SER (Switzerland).
Individuals from the CMS collaboration have received support from the Marie-Curie programme and the European Research
Council and EPLANET (European Union); the Leventis Foundation; the A. P. Sloan
Foundation; the Alexander von Humboldt Foundation; the Belgian Federal Science Policy Office;
the Fonds pour la Formation \`a la Recherche dans l'Industrie et dans l'Agriculture (FRIABelgium);
the Agentschap voor Innovatie door Wetenschap en Technologie (IWT-Belgium); the
Ministry of Education, Youth and Sports (MEYS) of the Czech Republic; the Council of Science
and Industrial Research, India; the HOMING PLUS programme of Foundation for Polish
Science, cofinanced from European Union, Regional Development Fund; the Compagnia di
San Paolo (Torino); the Consorzio per la Fisica (Trieste); MIUR project 20108T4XTM (Italy); the
Thalis and Aristeia programmes cofinanced by EU-ESF and the Greek NSRF; and the National
Priorities Research Program by Qatar National Research Fund.
Individual groups or members of the LHCb collaboration have received support from 
EPLANET, Marie Sk\l{}odowska-Curie Actions and ERC (European Union), 
Conseil g\'{e}n\'{e}ral de Haute-Savoie, 
Labex ENIGMASS and OCEVU, 
R\'{e}gion Auvergne (France), 
RFBR (Russia), 
XuntaGal and GENCAT (Spain), 
Royal Society and Royal Commission for the Exhibition of 1851 (UK).
LHCb is also thankful for the computing resources and the
access to software R\&D tools provided by Yandex LLC (Russia). 
The CMS and LHCb collaborations are indebted to the communities behind the multiple open source
software packages on which they depend.

\paragraph{Author Contributions} 
All authors have contributed to the publication, being variously involved in the design and the construction of the detectors, in writing 
software, calibrating sub-systems, operating the detectors and acquiring data and finally analysing the processed data.

\paragraph{Author Information} 
Reprints and permissions information is available at \nolinkurl{www.nature.com/reprints}.
The authors declare no competing financial interests.
Correspondence and requests for materials should be addressed to \nolinkurl{cms-publication-committee-chair@cern.ch} and 
to \nolinkurl{lhcb-editorial-board-chair@cern.ch}.

%% file: methods_extended.tex
\section*{Methods}

\input{experimental_setups}

\input{analyses}

%% file: experimental_setups.tex
\paragraph{Experimental Setup} 
At the Large Hadron Collider (LHC), two counter-rotating beams of protons, contained and guided by superconducting magnets spaced around a 27\,km circular tunnel, 
located approximately 100\m underground near Geneva, Switzerland, are brought into collision at four interaction points (IPs). 
The study presented in this Letter uses data collected at energies of 3.5\tev per beam in 2011 and 4\tev per beam in 2012 
by the CMS and LHCb experiments located at two of these IPs. 

The CMS and LHCb detectors are both designed to look for phenomena beyond the \sm (BSM), but using complementary strategies.
The CMS  detector~\cite{Chatrchyan:2008aa}, shown in Extended Data Fig.~\ref{fig:CMS_event_display}, is optimised to search for yet unknown heavy 
particles, with masses ranging from 100\gevcc to a few \tevcc, 
which, if observed, would be a direct manifestation of BSM phenomena.
Since many of the hypothesised new particles can decay into particles containing $b$ quarks or into muons, CMS is able to detect efficiently 
and study \Bd (5280\mevcc) and \Bs (5367\mevcc) mesons 
decaying to two muons  even though it is designed  to search for particles with much larger masses.
The CMS detector covers a very large range of angles and momenta to reconstruct high-mass states efficiently. 
To that extent, it employs a 13\,m long, 6\,m diameter superconducting solenoid magnet, operated at a field of 3.8\,T, 
centred on the IP with its axis along the beam direction and covering both hemispheres. 
A series of silicon tracking layers, consisting of silicon pixel detectors near the beam and
silicon strips farther out,
organised in  concentric cylinders around the beam, extending to a radius of 1.1\,m and terminated on each end by planar detectors (disks) 
perpendicular to the beam, measures the momentum, angles,  and position of charged particles emerging from the collisions. 
Tracking coverage starts from the direction perpendicular to the beam and extends to within 220\,mrad  from it
on both sides of the IP. The inner three cylinders and disks extending 
from 4.3 to 10.7\,cm in radius transverse to the beam are arrays of $100 \times 150\,\mu$m$^2$  silicon pixels, which can distinguish the displacement 
of the $b$-hadron decays from the primary vertex of the collision. The silicon strips, covering radii from 25\,cm to approximately 110\,cm,  have pitches ranging from 80 to 183 $\mu$m.    
The impact parameter is measured with a precision of 10$\rm \,\mu$m for transverse momenta of 100\gevc and 20$\rm \,\mu$m for 10\gevc.
The momentum resolution, provided mainly by the silicon strips, changes with the angle relative to the beam direction,  resulting in a mass resolution 
for \Bsdmumu decays that varies from 32\mevcc for \Bsd mesons produced perpendicularly to the proton beams to 75\mevcc for those produced at small angles relative to the beam direction.
After the tracking system, at a  greater distance from the IP, there is a calorimeter that stops (absorbs) all particles except muons and measures their energies. The calorimeter
consists of an  electromagnetic section followed by a hadronic section.
Muons are identified by their ability to penetrate the calorimeter and the steel return yoke of the solenoid magnet and to produce signals in gas-ionisation particle detectors located in 
compartments within the steel yoke.
The CMS detector has no capability to discriminate between charged hadron 
species,  pions, kaons, or protons, that is effective at the typical particle momenta in this analysis.

The primary commitment of the LHCb collaboration is the study of particle-antiparticle asymmetries and of rare decays of particles containing $b$ and $c$ quarks.
LHCb aims at detecting BSM particles indirectly by measuring their effect on $b$-hadron properties for which precise SM predictions exist.
The production cross section of $b$ hadrons at the LHC is particularly large at small angles relative to the colliding beams.
The small-angle region also provides advantages for the detection and reconstruction of a wide range of their decays.
The LHCb experiment~\cite{Alves:2008zz}, shown in Extended Data Fig.~\ref{fig:LHCb_event_display}, instruments the angular interval from 10 to 300\,mrad with 
respect to the beam direction on one side of the interaction region.
Its detectors are designed  to 
reconstruct efficiently a wide range of $b$-hadron decays, resulting in charged pions and kaons, protons, muons, electrons, and photons in the final state.
The detector includes a high-precision tracking system consisting of a silicon strip vertex detector, a large-area silicon strip detector located upstream of a dipole magnet 
characterised by a field integral of 4\unit{T\cdot m}, and three stations of silicon strip detectors and straw drift tubes downstream of the magnet.
The vertex detector has sufficient spatial resolution to distinguish the slight displacement of  the  weakly decaying $b$ hadron from the the primary production vertex where  the 
two protons collided and produced it.
The tracking detectors upstream and downstream of the dipole magnet measure the momenta of charged particles.
The combined tracking system provides a momentum measurement with an uncertainty that varies from 0.4\% at 5\gevc to 0.6\% at 100\gevc.
This results in an invariant-mass resolution of 25\mevcc for \Bsd mesons decaying to two muons that is nearly independent of the angle with respect to the beam.
The impact parameter resolution is smaller than 20\,${\rm \mu m}$ for particle tracks with large transverse momentum.
Different types of charged hadrons are distinguished by information from two ring-imaging Cherenkov detectors.
Photon, electron, and hadron candidates are identified by calorimeters.
Muons are identified by a system composed of alternating layers of iron and multiwire proportional chambers.

Neither CMS nor LHCb records all the interactions occurring at its IP because the data storage and analysis costs would be prohibitive. 
Since most of the interactions are reasonably well characterised  (and can be further studied by recording only a small sample of them) 
specific event filters (known as triggers) select the rare processes that are of interest to the experiments.
Both CMS and LHCb implement triggers that specifically select events containing two muons.
The triggers of both experiments have a hardware stage, based on information from the calorimeter and muon systems, followed by a software stage, 
consisting of  a large computing cluster that uses all the information from the detector, including the tracking, to make the final selection of 
events to be recorded for subsequent analysis. Since CMS is designed to look 
for much heavier objects than \Bsd mesons, it selects events that contain muons with higher transverse momenta than those selected by LHCb. 
This eliminates many of the \Bsd decays while permitting CMS to run
at a  higher proton-proton collision rate to look for the more rare massive particles. 
Thus CMS runs at higher collision rate but with lower efficiency than LHCb for \Bsd mesons decaying to two muons. 
The overall sensitivity to these decays turns out to be similar in the two experiments.

CMS and LHCb are not the only collaborations to have searched for \Bsmumu and \Bdmumu decays. Over three decades, a total of eleven collaborations have 
taken part in this search\cite{PDG2012}, as illustrated by Extended Data Fig.~\ref{fig:History}. This plot gathers the results from 
CLEO~\cite{Giles:1984yg,Avery:1987cv,Avery:1989qi,Ammar:1993ez,Bergfeld:2000ui},
ARGUS~\cite{Albrecht:1987rj},
UA1~\cite{Albajar:1988iq,Albajar:1991ct},
CDF~\cite{Abe:1996et,Abe:1998ah,Acosta:2004xj,Abulencia:2005pw,Aaltonen:2011fi,Aaltonen:2013as},
L3~\cite{Acciarri:1996us},
D\O~\cite{Abbott:1998hc,Abazov:2004dj,Abazov:2007iy,Abazov:2010fs,Abazov:2013wjb},
Belle~\cite{Chang:2003yy},
Babar~\cite{Aubert:2004gm,Aubert:2007hb},
LHCb~\cite{Aaij:2011rja,Aaij:2012ac,LHCb:2011ac,Aaij:2012nna,Aaij:2013aka},
CMS~\cite{Chatrchyan:2011kr,Chatrchyan:2012rga,Chatrchyan:2013bka},
and ATLAS~\cite{Aad:2012pn}. 

%% file: analyses.tex
\paragraph{Analysis description} 
The analysis techniques used to obtain the results presented in this Letter are very similar to those used to obtain
the individual result in each collaboration, described in more details in refs~\citen{LHCb-PAPER-2013-046,Chatrchyan:2013bka}.
Here only the main analysis steps are reviewed and the  changes used in the combined analysis are highlighted.  
Data samples for this analysis were collected by the two experiments in proton-proton 
collisions at a centre-of-mass energy of 7 and 8\tev during 2011 and 
2012, respectively. 
These samples correspond to a total integrated luminosity of 
$25$ and $3\invfb$ for the CMS and LHCb experiments, respectively, and represent their 
complete data sets from the first running period of the LHC. 

The trigger criteria 
were slightly different between the two experiments.
The large majority of events were triggered by requirements on one or both 
muons of the signal decay: the LHCb 
detector triggered on muons with transverse momentum $\pt>1.5\gevc$ while the CMS 
detector, because of its geometry and  higher instantaneous luminosity, triggered on two muons with $\pt > 
4 (3)\gevc$, for the leading (sub-leading) muon.

The data analysis procedures in the two experiments follow similar strategies. 
Pairs of high-quality oppositely charged particle tracks that have 
one of the expected  patterns of hits in the muon detectors are fitted to form a 
common vertex in three dimensions, which is required to be displaced from the primary interaction vertex (PV) and to 
have a small $\chi^2$ in the fit. The resulting \Bsd candidate is further required to point back
to the PV, for example to have a small impact parameter, consistent with zero, with respect to it.
The final classification of data events is done in categories of the response of a multivariate discriminant (MVA) combining information from the 
kinematics and vertex topology of the events. The type of MVA used is a boosted decision tree (BDT)~\cite{Breiman,Adaboost,Hocker:2007ht}.
The branching fractions are then obtained by a fit to the  dimuon invariant mass, 
\mmumu, of all categories simultaneously.  

The signals appear as peaks at the \Bs and \Bd masses in the invariant-mass distributions, observed over background events.
One of the components of the background is combinatorial in nature, as it is due to the random combinations of genuine muons. 
These produce a smooth dimuon mass distribution in the vicinity of the \Bs and \Bd masses, estimated in the fit to the data by extrapolation 
from the sidebands of the invariant-mass distribution. 
In addition to the combinatorial background, certain specific $b$-hadron 
decays can mimic the signal or contribute to the background in its vicinity. 
In particular, the semi-leptonic decays \bdpimunu, \bskmunu, 
\lbpmunu, can have reconstructed masses that are near the signal if one of the hadrons
is misidentified as a muon, and is combined with a genuine muon.
Similarly the dimuon coming from the rare \bpiZeromumu and \bpiPlusmumu decays can 
also fake the signal.
All these background decays, when reconstructed as a dimuon final state,
have invariant masses that are lower than the masses of the \Bd and \Bs mesons,
because they are missing one of the original decay particles.
An exception is the decay \lbpmunu, which can also populate, with a smooth mass distribution,
higher-mass regions.
Furthermore, background due to misidentified hadronic two-body decays \bhhprime, where $h^{(\prime)} = 
\pi$ or $K$, is present when both hadrons are misidentified as muons. These misidentified decays produce 
an apparent dimuon invariant-mass peak close to the $B^0$ mass value. 
Such a peak can mimic a \bdmumu signal and is estimated from control channels and added to the fit.

The distributions of signal in the invariant mass and in the MVA discriminant
are derived from simulations with a detailed description of the detector response 
for CMS and are calibrated using exclusive two-body hadronic decays in data for LHCb.
The distributions for the backgrounds are obtained from simulation with the exception 
of the combinatorial background. 
The latter is obtained by interpolating from the data invariant-mass sidebands separately for 
each category, after the subtraction of the other background components.

To compute the signal branching fractions, the numbers of \Bs and \Bd mesons that are produced, 
as well as the numbers of those that have decayed into a dimuon pair, are needed.
The latter numbers are the raw results of this analysis, whereas the former need to be determined from measurements of one or more `normalisation' decay channels, which are
abundantly produced, have an absolute branching fraction that is already known with good precision, and that share characteristics with the signals, 
so that their trigger and selection efficiencies do not differ significantly.
Both experiments use the \bujpsik decay as a normalisation channel with $\BRof\bujpsikmm = (6.10\pm0.19)\times 10^{-5}$, 
and LHCb also uses the \bdkpi channel with $\BRof\bdkpi = (1.96\pm0.05)\times 10^{-5}$.
Both branching fraction values are taken from ref.~\citen{PDG2012}.
Hence, the \bsmumu branching fraction is expressed as a function of the number of signal events 
($N_{\bsmumu}$) in the data normalised to the numbers of \bujpsik and \bdkpi events: 
\begin{equation}
 \mathcal{B}(\bsmumu) = \frac{N_{\bsmumu}}{N_{\rm{norm.{}}}}
\times\frac{f_d}{f_s} \times \frac{\varepsilon_{\rm{norm.{}}}}{\varepsilon_{\bsmumu}} 
\times \mathcal{B}_{\rm{norm.{}}} = \alpha_{\rm{norm.{}}} \times N_{\bsmumu} ,
\label{eq:normalisation}
\end{equation}
where the `norm.{}' subscript refers to either of the normalisation channels.
The values of the normalisation parameter $\alpha_{\rm{norm.{}}}$ obtained by LHCb from the two normalisation channels are found in good agreement 
and their weighted average is used.
In this formula $\varepsilon$ indicates the total event detection efficiency including 
geometrical acceptance, trigger selection, reconstruction, and analysis selection for the corresponding decay. 
The \fdfs factor is the ratio of the probabilities for a $b$ quark to hadronise into a \Bd as compared 
to a \Bs meson; the probability to hadronise into a \Bu ($f_u$) is assumed to be equal to that into \Bd ($f_d$) on the basis of theoretical grounds, and this assumption is checked on data. 
The value of $\fdfs = 3.86\pm0.22$ measured by LHCb~\cite{LHCb-CONF-2013-011,LHCb-PAPER-2012-037,LHCb-PAPER-2011-018} is used in  
this analysis. As the value of \fdfs depends on the kinematic range of the considered particles, which differs between LHCb and CMS, 
CMS checked this observable with the decays \BsJpsiPhi and \BuJpsiK  
within its acceptance, finding a consistent value. An additional 
systematic uncertainty of $5\%$ was assigned to \fdfs to account for the extrapolation of the LHCb result to the CMS
acceptance. 
An analogous formula to that in equation~\eqref{eq:normalisation} holds for the normalisation of the 
\bdmumu decay, with the notable difference that the \fdfs factor is replaced by $\fdfuinline=1$. 

The antiparticle \Bdb (\Bsb) and the particle \Bd (\Bs) can both decay into two muons  and no attempt is made in this 
analysis to determine whether the antiparticle or particle was produced (untagged method). 
However, the \Bd and \Bs particles are known to oscillate, that is to transform continuously into their antiparticles and vice 
versa. 
Therefore, a quantum superposition of particle and antiparticle states propagates in the laboratory before decaying. 
This superposition can be described by two `mass eigenstates', which are symmetric and anti-symmetric in the charge-parity (CP) quantum number, 
and have slightly different masses. 
In the \sm, the heavy eigenstate can decay into two muons, whereas the light eigenstate cannot
without violating the CP quantum number conservation.
In BSM models, this is not necessarily the case. 
In addition to their masses, the two eigenstates of the \Bs system also differ in their lifetime values~\cite{PDG2012}.
The lifetimes of the light and heavy eigenstates are also different from the average \Bs lifetime, which is used 
by CMS and LHCb in the simulations of signal decays. 
Since the information on the displacement of the secondary decay with respect to the 
PV is used as a discriminant against combinatorial background in the analysis, 
the efficiency versus lifetime has a model-dependent bias~\cite{DeBruyn:2012wk} that must be removed. 
This bias is estimated assuming \sm dynamics. 
Owing to the smaller difference between the lifetime of its heavy and light mass eigenstates, 
no correction is required for the \Bd decay mode.

Detector simulations are needed by both CMS and LHCb.
CMS relies on simulated events to determine resolutions and trigger and reconstruction efficiencies,
and to provide the signal sample for training the BDT.
The dimuon mass resolution given by the simulation is validated using data on  $J/\psi$, $\Upsilon$, and
$Z$-boson decays to two muons.
The tracking and trigger efficiencies obtained from the simulation are checked 
using special control samples from data.
The LHCb analysis is designed to minimise the impact of discrepancies between simulations and data. 
The mass resolution is measured with data. The distribution of the BDT for the signal and 
for the background is also calibrated with data using control channels and mass sidebands.
The efficiency ratio for the trigger is also largely determined from data.
The simulations are used to determine the efficiency ratios of selection and reconstruction processes 
between signal and normalisation channels.
As for the overall detector simulation, each experiment has a  team dedicated to making the simulations
as complete and realistic as possible.  The simulated data are constantly being compared to the actual data. 
Agreement between simulation and data  in both experiments is quite good, often extending well 
beyond the cores of distributions. Differences occur because, 
for example, of incomplete description of the material of the detectors, approximations made to 
keep the computer time manageable, residual uncertainties in calibration and alignment, and discrepancies or
limitations in the underlying theory and experimental data used to model the relevant collisions and decays.
Small differences between simulation and data that are known to have an  impact on  the result are
treated either by reweighting the simulations to match the data or by assigning appropriate systematic
uncertainties.

Small changes are made to the analysis procedure with respect to 
refs~\citen{LHCb-PAPER-2013-046,Chatrchyan:2013bka} in order to achieve a consistent combination between 
the two experiments. 
In the LHCb analysis, the \lbpmunu background component, which was not included in the fit for the previous result but
whose effect was accounted for as an additional systematic uncertainty, is now included in the standard fit.
The following modifications are made to the CMS analysis: the \lbpmunu 
branching fraction is updated to a more recent prediction~\cite{Khodjamirian:2011jp,LHCB-PAPER-2014-003} of $\BRof\lbpmunu = (4.94\pm2.19)\times 10^{-4}$; 
the phase space model of the decay \lbpmunu 
is changed to a more appropriate semi-leptonic decay model~\cite{Khodjamirian:2011jp}; 
and the decay time bias correction for the \Bs, previously absent from the analysis, is now calculated and applied with a different correction for each category of 
the multivariate discriminant.

These modifications result in changes in the individual results of each experiment.
The modified CMS analysis, applied on the CMS data, yields 
\begin{eqnarray}
\BRof\Bsmumu  =  \CMSBsmumuMeasureMod  \quad\text{ and }\quad   \BRof\Bdmumu  = \CMSBdmumuMeasureMod,
\end{eqnarray}
while the LHCb results change to  
\begin{eqnarray}
\BRof\Bsmumu  = \LHCbBsmumuMeasureMod \quad\text{ and }\quad  \BRof\Bdmumu =  \LHCbBdmumuMeasureMod.
\end{eqnarray}
These results are only slightly different from the published ones and are in agreement with each other. 

\paragraph{Simultaneous fit}

The goal of the analysis presented in this Letter is to combine the full data sets of the 
two experiments to reduce the uncertainties on the branching fractions of the signal decays
obtained from the individual determinations. 
A simultaneous unbinned extended maximum likelihood fit is performed to the data of 
the two experiments,
using the invariant-mass distributions of all 20 MVA discriminant categories of both experiments.
The invariant-mass distributions are defined in the dimuon mass ranges $m_{\mu^+ \mu^-} \in [4.9, 5.9]\gevcc$
and  $[4.9, 6.0]\gevcc$ for the CMS and LHCb experiments, respectively.
The branching fractions of the signal decays, 
the hadronisation fraction ratio \fdfs, and the branching fraction of the normalisation channel \BuJpsiK 
are treated as common parameters.
\tblue{The value of the  \BuJpsiK branching fraction is the combination of  
results from five different experiments~\cite{PDG2012}, taking advantage of all their data 
to achieve the most precise input parameters for this analysis.} 
The combined fit takes advantage of the larger data sample and proper treatment of the correlations between
the individual measurements to increase the precision and reliability of the result, respectively.

Fit parameters, other than those of primary physics interest, whose limited knowledge affects the results, are  called `nuisance parameters'.
In particular, systematic uncertainties are modelled by introducing nuisance parameters into the statistical model 
and allowing them to vary in the fit; those for which additional knowledge is present are constrained using Gaussian distributions. 
The mean and standard deviation of these distributions are set 
to the central value and uncertainty obtained either from other measurements or from control channels. 
The statistical component of the final uncertainty on the branching fractions is obtained by repeating the fit 
after fixing all of the constrained nuisance parameters to their best fitted values.  
The systematic component is then calculated by subtracting in quadrature the statistical component from the total uncertainty.
In addition to the free fit, a two-dimensional likelihood ratio scan in the plane  \BRof\Bdmumu versus \BRof\Bsmumu is 
performed. 

\paragraph{Feldman--Cousins Confidence Interval}
The Feldman--Cousins likelihood ratio ordering procedure~\cite{Feldman:1997qc} is a unified frequentist method 
to construct single- and double-sided confidence intervals for parameters of a given model adapted to the data.
It provides a natural transition between single-sided confidence intervals, used to define upper or lower limits, and double-sided ones.
Since the single-experiment results~\cite{LHCb-PAPER-2013-046,Chatrchyan:2013bka} showed that the \Bdmumu 
signal is at the edge of the probability region customarily used to assert statistically
significant evidence for a result, a Feldman--Cousins procedure is performed.
This allows a more reliable determination of the confidence interval and significance of this signal without the assumptions required for the use of Wilks' theorem. 
In addition, a prescription for the treatment of nuisance parameters has to be chosen because scanning the whole parameter space in the presence of more than a few 
parameters is computationally too intensive. 
In this case the procedure described by the ATLAS and CMS Higgs combination group~\cite{Higgs-Comb} is adopted.
For each point of the space of the relevant parameters, the nuisance parameters 
are fixed to their best value estimated by the mean of a maximum likelihood fit to the data with the value of \BRof\Bdmumu fixed and all nuisance 
parameters profiled with Gaussian penalties.
Sampling distributions are constructed for each tested point of the parameter of interest    
by generating simulated experiments and performing maximum likelihood fits  in which the Gaussian mean values of the external constraints on the nuisance parameters
are randomised around the best-fit values for the nuisance parameters used to generate the simulated experiments.
The sampling distribution is constructed from the distribution of the negative log-likelihood ratio evaluated on the simulated experiments by performing one likelihood fit
in which the value of \BRof\Bdmumu is free to float and another with the \BRof\Bdmumu fixed to the tested point value. This sampling distribution is then converted to a confidence 
level by evaluating the fraction
of simulated experiments entries with a value for the negative log-likelihood ratio greater than or equal to the value observed in the data for each tested point. The results of this
procedure are shown in Extended Data Fig.~\ref{fig:fc_results_lhcb}.

%% file: ref2.tex
\ifx\mcitethebibliography\mciteundefinedmacro
\PackageError{LHCb.bst}{mciteplus.sty has not been loaded}
{This bibstyle requires the use of the mciteplus package.}\fi
\providecommand{\href}[2]{#2}

%% file: CMS_AL.tex
\begin{flushleft}
\small
\noindent\textbf{The CMS Collaboration:}
V.~Khachatryan$^{1}$,
A.M.~Sirunyan$^{1}$,
A.~Tumasyan$^{1}$,
W.~Adam$^{2}$,
T.~Bergauer$^{2}$,
M.~Dragicevic$^{2}$,
J.~Er\"{o}$^{2}$,
M.~Friedl$^{2}$,
R.~Fr\"{u}hwirth$^{2,b}$,
V.M.~Ghete$^{2}$
% ...
\\[2ex]
$^{1}$~Yerevan Physics Institute, Yerevan, Armenia\\
$^{2}$~Institut f\"{u}r Hochenergiephysik der OeAW, Wien, Austria\\
% ...
\hrulefill\\[1ex]
\textit{\footnotesize
a~Deceased\\ 
b~Also~at~Vienna University of Technology, Vienna, Austria\\
c~Also~at~CERN, European Organization for Nuclear Research, Geneva, Switzerland\\
d~Also~at~Institut Pluridisciplinaire Hubert Curien, Universit\'{e} de Strasbourg, Universit\'{e} de Haute Alsace Mulhouse, CNRS/IN2P3, Strasbourg, France\\
e~Also~at~National Institute of Chemical Physics and Biophysics, Tallinn, Estonia\\
f~Also~at~Skobeltsyn Institute of Nuclear Physics, Lomonosov Moscow State University, Moscow, Russia\\
g~Also~at~Universidade Estadual de Campinas, Campinas, Brazil\\
h~Also~at~Laboratoire Leprince-Ringuet, Ecole Polytechnique, IN2P3-CNRS, Palaiseau, France\\
i~Also~at~Joint Institute for Nuclear Research, Dubna, Russia\\
j~Also~at~Suez University, Suez, Egypt\\
k~Also~at~Cairo University, Cairo, Egypt\\
l~Also~at~Fayoum University, El-Fayoum, Egypt\\
m~Also~at~Ain Shams University, Cairo, Egypt\\
n~Now~at~Sultan Qaboos University, Muscat, Oman\\
o~Also~at~Universit\'{e} de Haute Alsace, Mulhouse, France\\
p~Also~at~Brandenburg University of Technology, Cottbus, Germany\\
q~Also~at~Institute of Nuclear Research ATOMKI, Debrecen, Hungary\\
r~Also~at~E\"{o}tv\"{o}s Lor\'{a}nd University, Budapest, Hungary\\
s~Also~at~University of Debrecen, Debrecen, Hungary\\
t~Also~at~University of Visva-Bharati, Santiniketan, India\\
u~Now~at~King Abdulaziz University, Jeddah, Saudi Arabia\\
v~Also~at~University of Ruhuna, Matara, Sri Lanka\\
w~Also~at~Isfahan University of Technology, Isfahan, Iran\\
x~Also~at~University of Tehran, Department of Engineering Science, Tehran, Iran\\
y~Also~at~Plasma Physics Research Center, Science and Research Branch, Islamic Azad University, Tehran, Iran\\
z~Also~at~Universit\`{a} degli Studi di Siena, Siena, Italy\\
aa~Also~at~Centre National de la Recherche Scientifique (CNRS) - IN2P3, Paris, France\\
bb~Also~at~Purdue University, West Lafayette, USA\\
cc~Also~at~Universidad Michoacana de San Nicolas de Hidalgo, Morelia, Mexico\\
dd~Also~at~Institute for Nuclear Research, Moscow, Russia\\
ee~Also~at~St. Petersburg State Polytechnical University, St. Petersburg, Russia\\
ff~Also~at~California Institute of Technology, Pasadena, USA\\
gg~Also~at~Faculty of Physics, University of Belgrade, Belgrade, Serbia\\
hh~Also~at~Facolt\`{a} Ingegneria, Universit\`{a} di Roma, Roma, Italy\\
ii~Also~at~Scuola Normale e Sezione dell'INFN, Pisa, Italy\\
jj~Also~at~University of Athens, Athens, Greece\\
kk~Also~at~Paul Scherrer Institut, Villigen, Switzerland\\
ll~Also~at~Institute for Theoretical and Experimental Physics, Moscow, Russia\\
mm~Also~at~Albert Einstein Center for Fundamental Physics, Bern, Switzerland\\
nn~Also~at~Gaziosmanpasa University, Tokat, Turkey\\
oo~Also~at~Adiyaman University, Adiyaman, Turkey\\
pp~Also~at~Cag University, Mersin, Turkey\\
qq~Also~at~Anadolu University, Eskisehir, Turkey\\
rr~Also~at~Ozyegin University, Istanbul, Turkey\\
ss~Also~at~Izmir Institute of Technology, Izmir, Turkey\\
tt~Also~at~Necmettin Erbakan University, Konya, Turkey\\
uu~Also~at~Mimar Sinan University, Istanbul, Istanbul, Turkey\\
vv~Also~at~Marmara University, Istanbul, Turkey\\
ww~Also~at~Kafkas University, Kars, Turkey\\
xx~Also~at~Yildiz Technical University, Istanbul, Turkey\\
yy~Also~at~Rutherford Appleton Laboratory, Didcot, United Kingdom\\
zz~Also~at~School of Physics and Astronomy, University of Southampton, Southampton, United Kingdom\\
aaa~Also~at~University of Belgrade, Faculty of Physics and Vinca Institute of Nuclear Sciences, Belgrade, Serbia\\
bbb~Also~at~Argonne National Laboratory, Argonne, USA\\
ccc~Also~at~Erzincan University, Erzincan, Turkey\\
ddd~Also~at~Texas A\&M University at Qatar, Doha, Qatar\\
eee~Also~at~Kyungpook National University, Daegu, Korea\\
}
\end{flushleft}

%% file: LHCb_AL.tex
\begin{flushleft}
\small
\noindent\textbf{The LHCb Collaboration:}
I.~Bediaga$^{1}$,
J.M.~De~Miranda$^{1}$,
F.~Ferreira~Rodrigues$^{1}$,
A.~Gomes$^{1,m}$,
A.~Massafferri$^{1}$,
A.C.~dos~Reis$^{1}$,
A.B.~Rodrigues$^{1}$,
S.~Amato$^{2}$,
K.~Carvalho~Akiba$^{2}$,
L.~De~Paula$^{2}$
% ...
\\[2ex]
$^{1}$~Centro Brasileiro de Pesquisas F\'{i}sicas (CBPF), Rio de Janeiro, Brazil\\
$^{2}$~Universidade Federal do Rio de Janeiro (UFRJ), Rio de Janeiro, Brazil\\
% ...
\hrulefill\\[1ex]
\textit{\footnotesize
a~Deceased\\ 
b~Also~at~Universit\`{a} di Firenze, Firenze, Italy\\
c~Also~at~Universit\`{a} di Ferrara, Ferrara, Italy\\
d~Also~at~Universit\`{a} della Basilicata, Potenza, Italy\\
e~Also~at~Universit\`{a} di Modena e Reggio Emilia, Modena, Italy\\
f~Also~at~Universit\`{a} di Milano Bicocca, Milano, Italy\\
g~Also~at~LIFAELS, La Salle, Universitat Ramon Llull, Barcelona, Spain\\
h~Also~at~Universit\`{a} di Bologna, Bologna, Italy\\
i~Also~at~Universit\`{a} di Roma Tor Vergata, Roma, Italy\\
j~Also~at~Universit\`{a} di Genova, Genova, Italy\\
k~Also~at~Scuola Normale Superiore, Pisa, Italy\\
l~Also~at~Politecnico di Milano, Milano, Italy\\
m~Also~at~Universidade Federal do Tri\^{a}ngulo Mineiro (UFTM), Uberaba-MG, Brazil\\
n~Also~at~AGH - University of Science and Technology, Faculty of Computer Science, Electronics and Telecommunications, Krak\'{o}w, Poland\\
o~Also~at~Universit\`{a} di Padova, Padova, Italy\\
p~Also~at~Universit\`{a} di Cagliari, Cagliari, Italy\\
q~Also~at~Hanoi University of Science, Hanoi, Viet Nam\\
r~Also~at~Universit\`{a} di Bari, Bari, Italy\\
s~Also~at~Universit\`{a} degli Studi di Milano, Milano, Italy\\
t~Also~at~Universit\`{a} di Roma La Sapienza, Roma, Italy\\
u~Also~at~Universit\`{a} di Pisa, Pisa, Italy\\
v~Also~at~Universit\`{a} di Urbino, Urbino, Italy\\
w~Also~at~P.N. Lebedev Physical Institute, Russian Academy of Science (LPI RAS), Moscow, Russia\\
}
\end{flushleft}

%% file: extended_data.tex
%
\newlength{\myVerticalShift}
\setlength{\myVerticalShift}{30mm}
\changepage{\myVerticalShift}{}{}{}{}{-\myVerticalShift}{}{}{}
%
%
\begin{extended}[p]
\centering
\includegraphics[width=\textwidth]{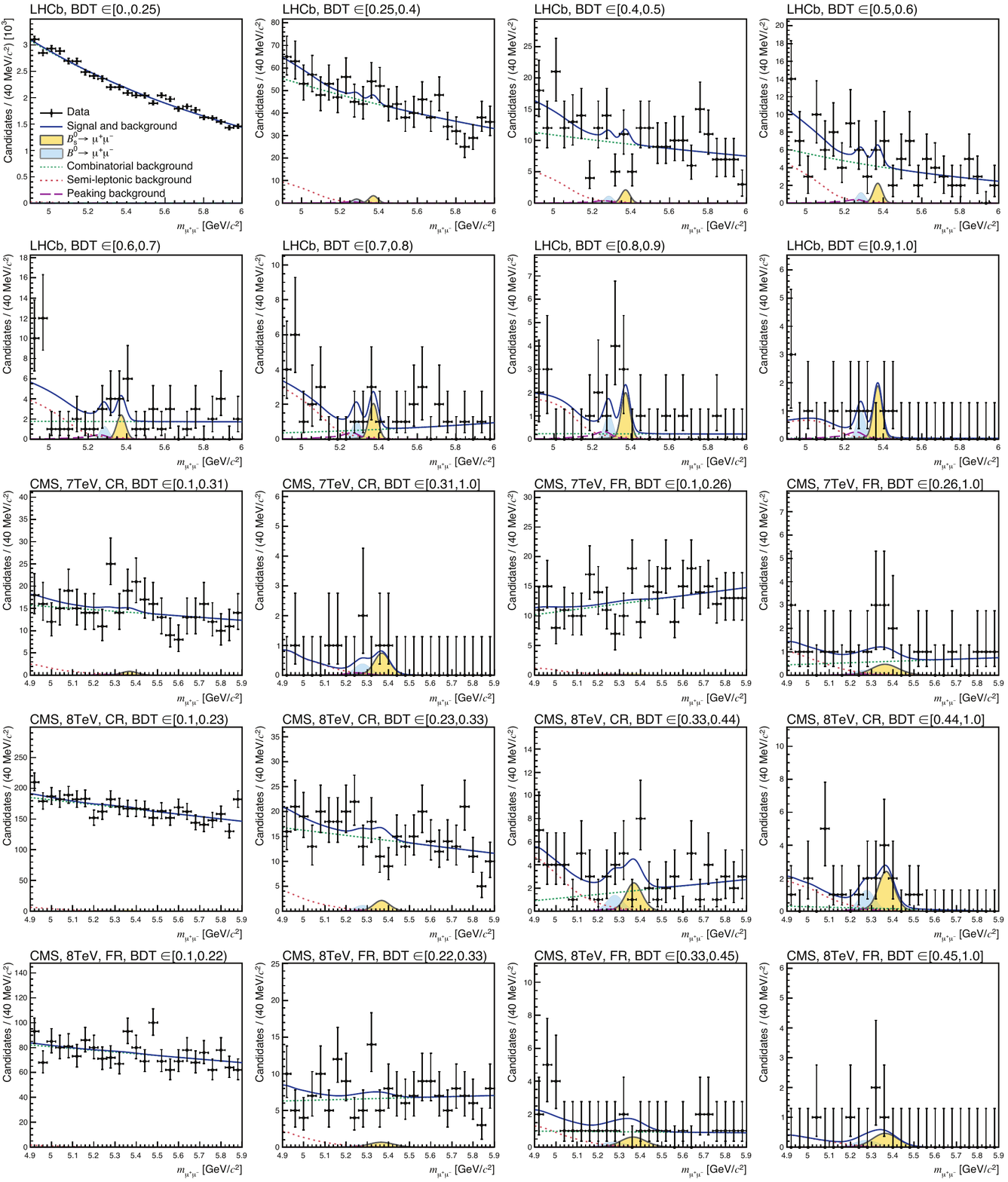}
\vspace{-25mm}
\caption{\textbf{Distribution of the dimuon invariant mass $m_{\mu^+\mu^-}$ in each of the 20 categories.}
Superimposed on the data points in black are the combined fit (solid blue) and its components: the \Bs (yellow shaded) and \Bd (light-blue shaded) signal components; 
the combinatorial background (dash-dotted green); 
the sum of the semi-leptonic backgrounds (dotted salmon); and the peaking backgrounds (dashed violet).
The categories are defined by the range of BDT values for LHCb, and for CMS, by centre-of-mass energy, by the region of the detector in which the muons are detected, and by the range of BDT values. 
Categories for which both muons are detected in the central region of the CMS detector are denoted with CR, those for which at least one muon was detected into the forward 
region with FR.
\label{fig:AllBins}}
\end{extended}
%
%
\begin{extended}[p]
\centering
\includegraphics[width=\textwidth]{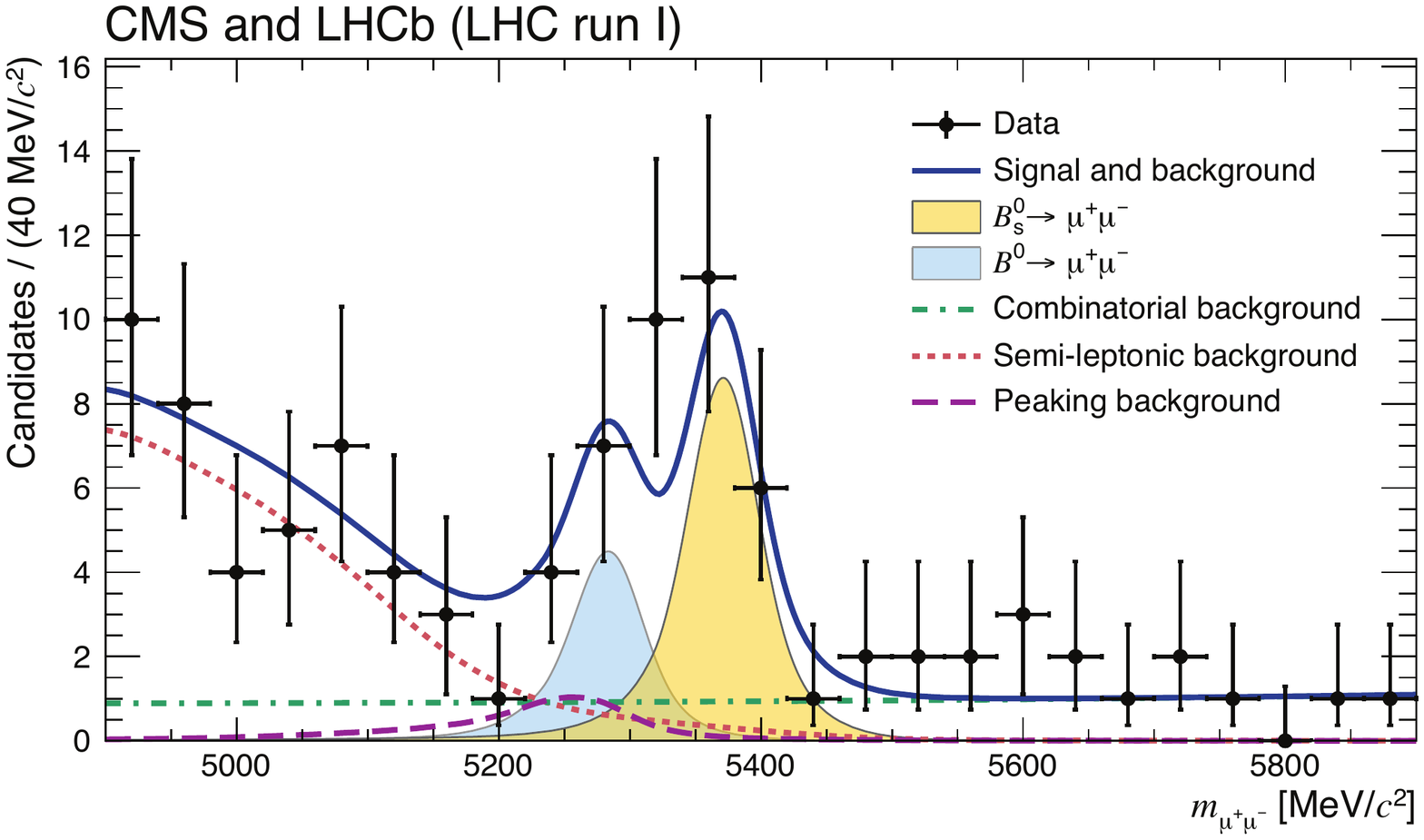}
\vspace{-70mm}
\caption{\textbf{Distribution of the dimuon invariant mass $m_{\mu^+\mu^-}$ for the best six categories.} 
Categories are ranked according to values of ${\rm S}/{\rm (S+B)}$ where S and B are the numbers of signal events expected assuming the \sm rates and background 
events under the \Bs peak for a given category, respectively.
The mass distribution for the six highest-ranking categories, three per experiment, is shown.
Superimposed on the data points in black are the combined full fit (solid blue) and its components: 
the \Bs (yellow shaded) and \Bd (light-blue shaded) signal components; the combinatorial background (dash-dotted green); 
the sum of the semi-leptonic backgrounds (dotted salmon); and the peaking backgrounds (dashed violet).\label{fig:MassBestBins}}
\end{extended}
%
%
\begin{extended}[p]
\centering
\includegraphics[width=\textwidth]{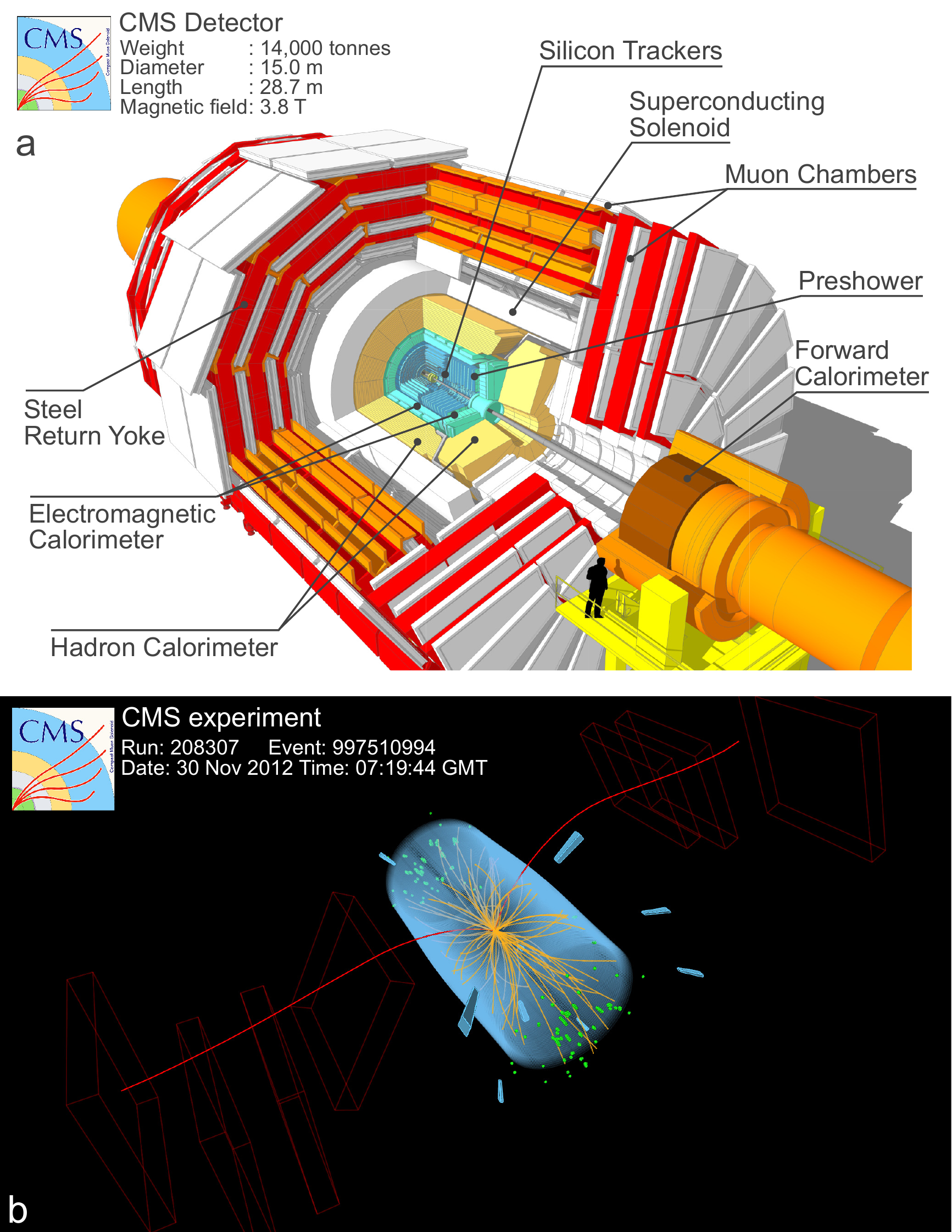}
\caption{\textbf{\boldmath Schematic of the CMS detector and event display for a candidate \Bsmumu decay at CMS.}
\textbf{\textsf{a}}, The CMS detector and its components; see ref.~\citen{Chatrchyan:2008aa} for details.
\textbf{\textsf{b}}, A candidate \Bsmumu decay produced in proton-proton collisions at 8\tev in 2012 and recorded in the CMS detector.  
The red arched curves represent the trajectories of the muons from the \Bs decay candidate.
\label{fig:CMS_event_display}}
\end{extended}
%
%
\begin{extended}[p]
\centering
\includegraphics[width=\textwidth]{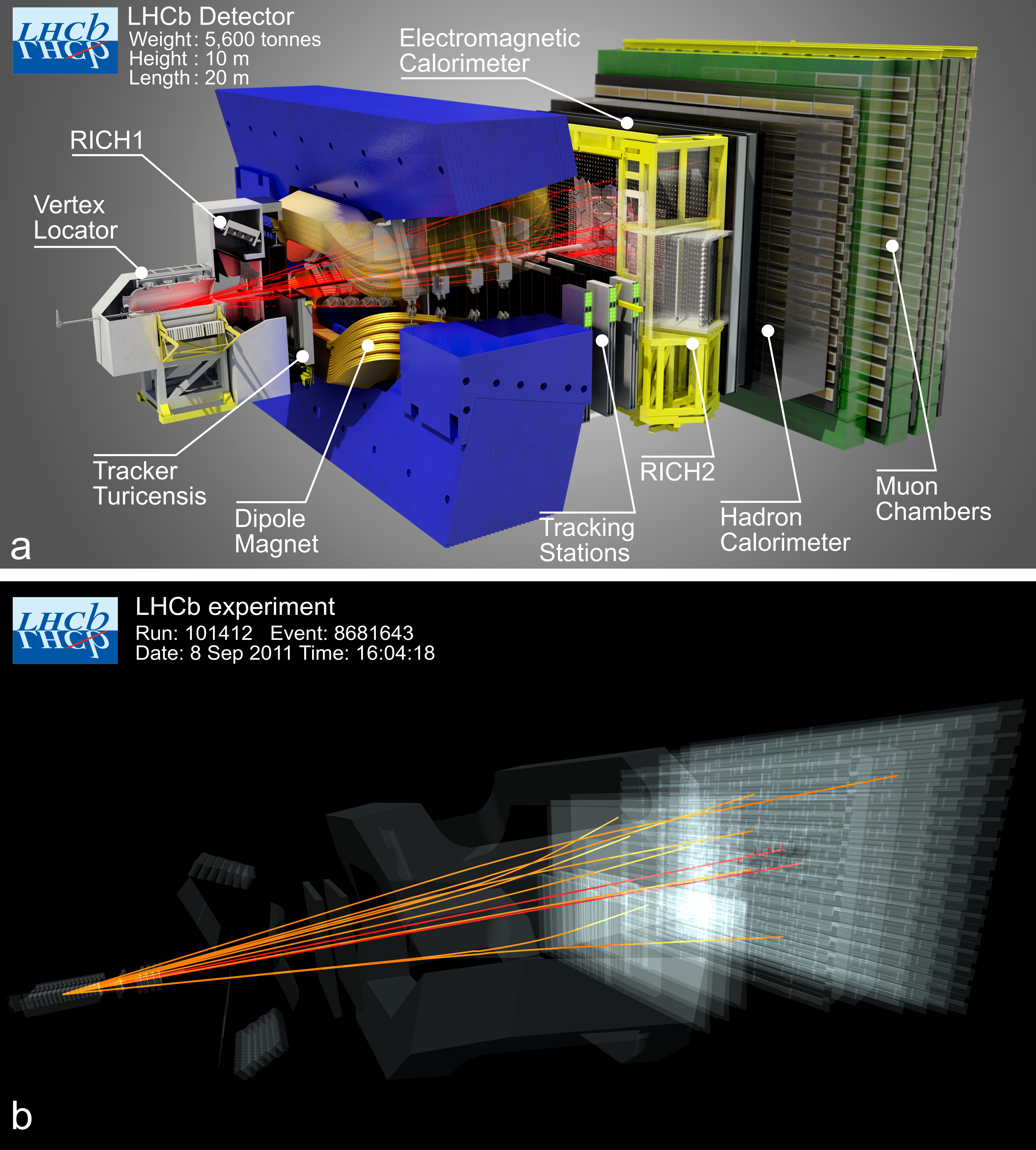}
\caption{\textbf{\boldmath Schematic of the LHCb detector and event display for a candidate \Bsmumu decay at LHCb.}
\textbf{\textsf{a}}, The LHCb detector and its components; see ref.~\citen{Alves:2008zz} for details.
\textbf{\textsf{b}}, A candidate \Bsmumu decay produced in proton-proton collisions at 7\tev in 2011 and recorded in the LHCb detector. 
The proton-proton collision occurs on the left-hand side, at the origin of the trajectories depicted with the orange curves. 
The red curves represent the trajectories of the muons from the \Bs candidate decay.
\label{fig:LHCb_event_display}}
\end{extended}
%
%
\begin{extended}[p]
\centering
\includegraphics[width=\textwidth]{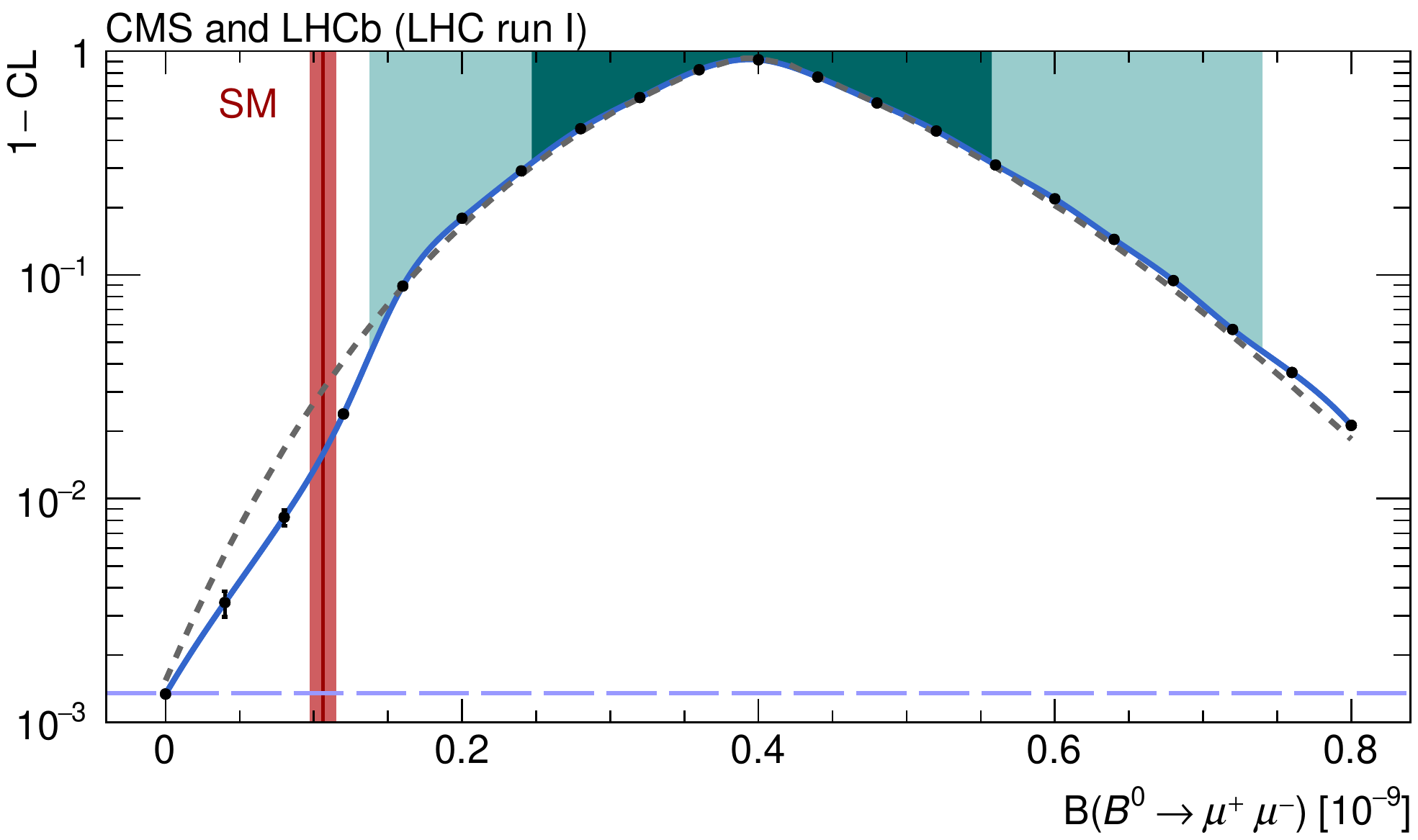}
\caption{\textbf{\boldmath Confidence level as a function of the \BRof{\Bdmumu} hypothesis.}   
Value of $1-\mathrm{CL}$, where CL is the confidence level obtained with the Feldman--Cousins procedure, 
as a function of \BRof{\Bdmumu} is shown in logarithmic scale. 
The points mark the computed $1-\mathrm{CL}$ values and the curve is their spline
interpolation.
The dark and light (cyan) areas define the two-sided $\pm1\sigma$ and $\pm2\sigma$ confidence intervals for the branching fraction, 
while the dashed horizontal line defines the confidence level for the  $3\sigma$ one-sided interval.
The dashed (grey) curve shows the $1-\mathrm{CL}$ values computed from the one-dimensional $-2\Delta\textrm{ln}L$ test statistic using Wilks' theorem.
Deviations between these confidence level values and those from the Feldman--Cousins procedure~\cite{Feldman:1997qc} illustrate the degree of approximation 
implied by the asymptotic assumptions inherent to Wilks' theorem~\cite{Wilks}.
\label{fig:fc_results_lhcb}}
\end{extended}
%
%
\begin{extended}[p]
\centering
\includegraphics[width=\textwidth ]{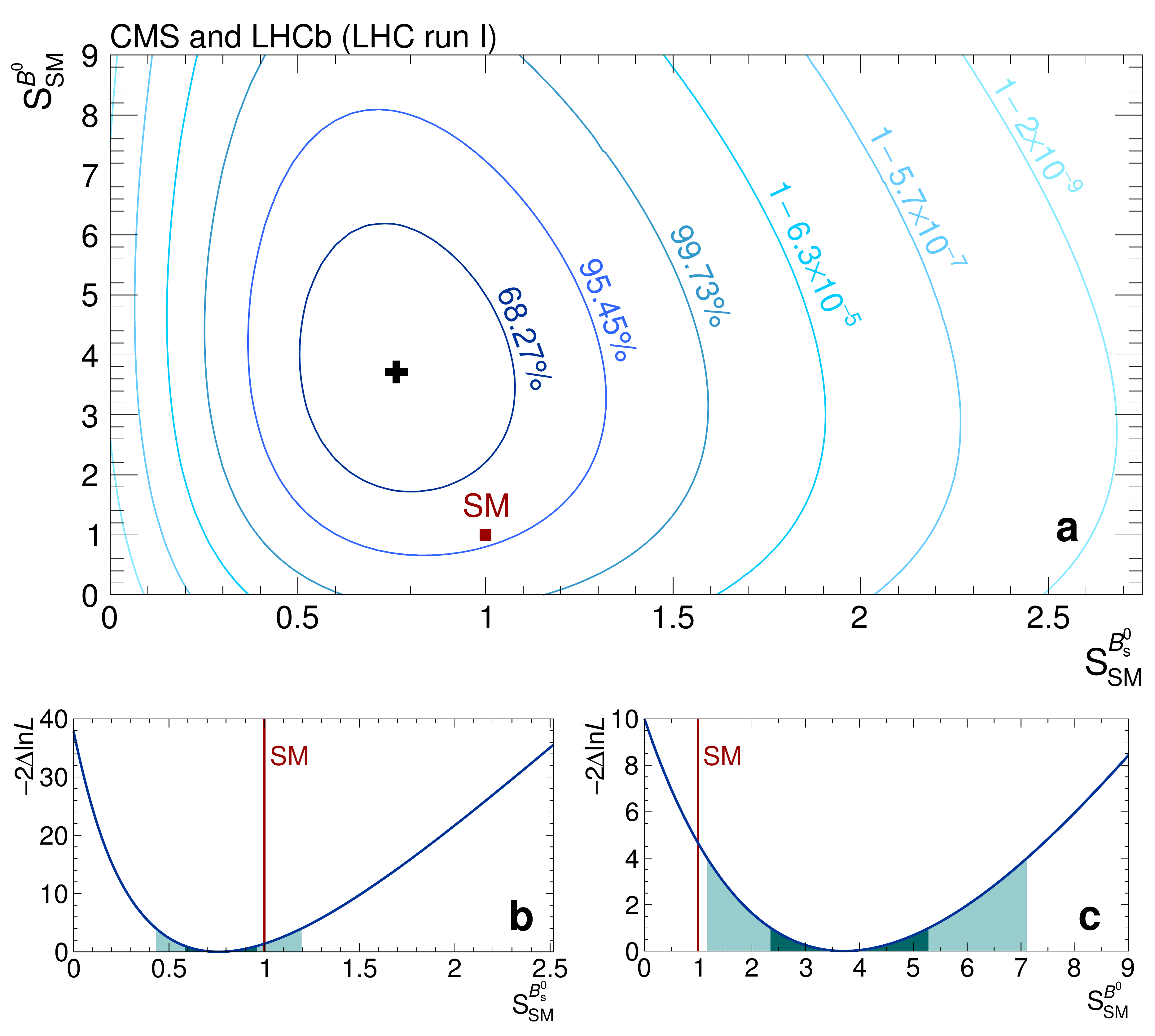}
\caption{\textbf{\boldmath Likelihood contours for the ratios of the branching fractions with respect to their \sm prediction, in the \SBd versus \SBs plane.}
\textbf{\textsf{a}}, The (black) cross marks the central value returned by the fit. 
The \sm point is shown as the (red) square located, by construction, at $\SBd = \SBs=1$.
Each contour encloses a region approximately corresponding to the reported confidence level.
The \sm branching fractions are assumed uncorrelated to each other, and their uncertainties are accounted for in the likelihood contours.
\textbf{\textsf{b}}, \textbf{\textsf{c}}, Variations of the test statistic $-2\Delta\textrm{ln}L$ for \SBs and \SBd are shown in 
\textbf{\textsf{b}} and \textbf{\textsf{c}}, respectively.
The \sm is represented by the (red) vertical lines.
The dark and light (cyan) areas define the $\pm1\sigma$ and $\pm2\sigma$ confidence intervals, respectively. 
\label{fig:BFoverSM}}
\end{extended}
\begin{extended}[p]
\centering
\includegraphics[width=\textwidth]{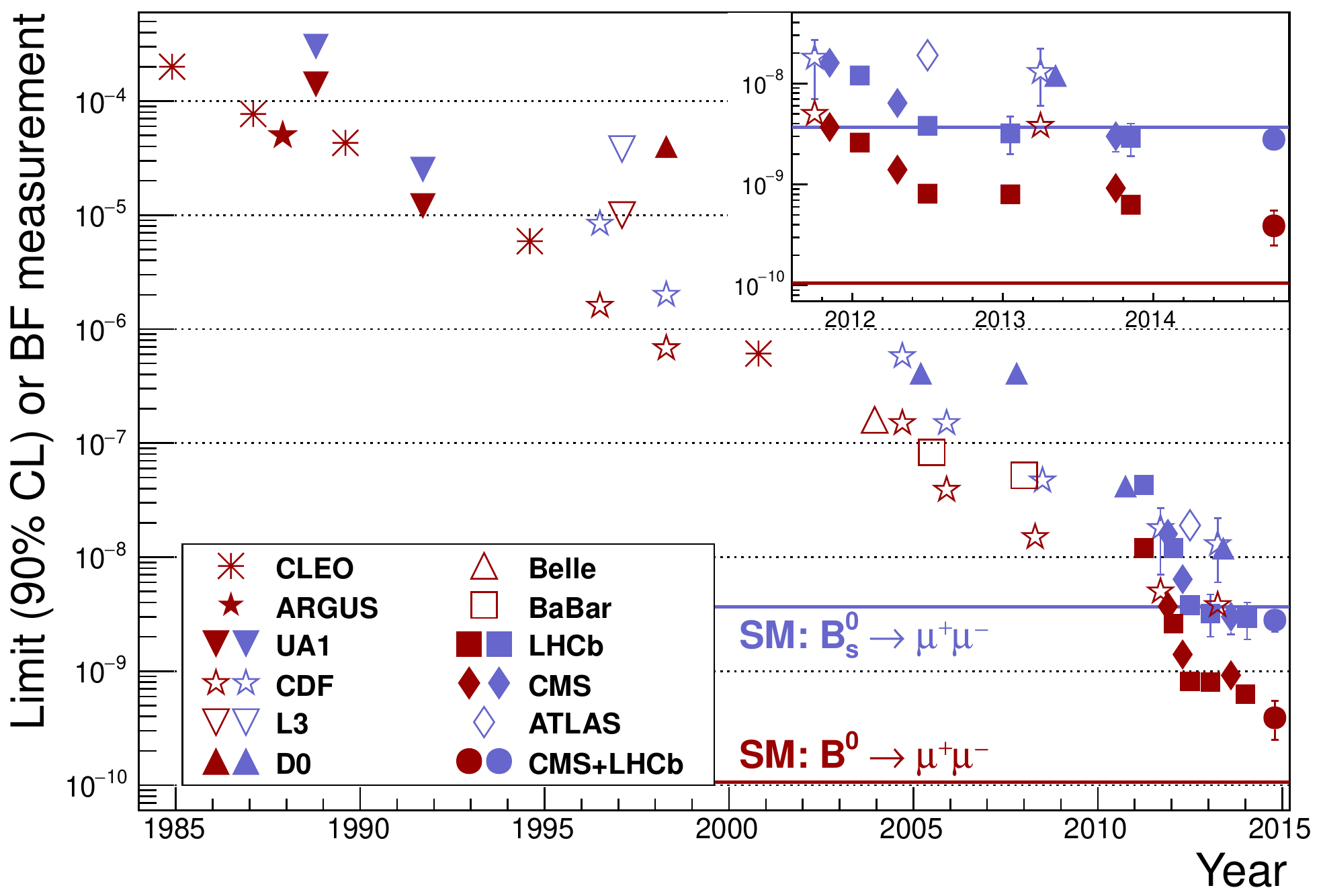}
\caption{\textbf{\boldmath Search for the \Bsmumu and \Bdmumu decays,
reported by 11 experiments spanning more than three decades, and by the present results.}
Markers without error bars denote upper limits on the branching fractions at 90\% confidence level, 
while measurements are denoted with errors bars delimiting 68\% confidence intervals.
The horizontal lines represent the SM predictions for the \Bsmumu and \Bdmumu branching fractions~\cite{Bobeth:2013uxa};
the blue (red) lines and markers relate to the \Bsmumu (\Bdmumu) decay.
Data (see key) are from refs%
~\citen{Giles:1984yg,Avery:1987cv,Avery:1989qi,Ammar:1993ez,Bergfeld:2000ui,Albrecht:1987rj
,Albajar:1988iq,Albajar:1991ct
,Abe:1996et,Abe:1998ah,Acosta:2004xj,Abulencia:2005pw,Aaltonen:2011fi,Aaltonen:2013as
,Acciarri:1996us
,Abbott:1998hc,Abazov:2004dj,Abazov:2007iy,Abazov:2010fs,Abazov:2013wjb
,Chang:2003yy
,Aubert:2004gm,Aubert:2007hb
,Aaij:2011rja,Aaij:2012ac,LHCb:2011ac,Aaij:2012nna,Aaij:2013aka
,Chatrchyan:2011kr,Chatrchyan:2012rga,Chatrchyan:2013bka
,Aad:2012pn}
; for details see Methods. Inset, magnified view  of the last period in time.
}\label{fig:History}
\end{extended}
\FloatBarrier